\documentclass[12pt,preprint]{aastex}

\begin{document}

\title{State Transitions in Bright Galactic X-ray Binaries: Luminosities Span by Two Orders of Magnitude}

\author{Wenfei Yu and Zhen Yan\altaffilmark{1}}
\affil{Key Laboratory for Research in Galaxies and Cosmology and Shanghai Astronomical Observatory, 80 Nandan Road, Shanghai 200030, China. E-mail: wenfei@shao.ac.cn}
\altaffiltext{1}{also Graduate School of the Chinese Academy of Sciences}

\begin{abstract}
Using X-ray monitoring observations with the All Sky Monitor (ASM) on board the {\it Rossi X-ray Timing Explorer (RXTE)} and the Burst Alert Telescope (BAT) on board the {\it Swift}, we are able to study the spectral state transitions occurred in about 20 bright persistent and transient black hole and neutron star binaries. We have confirmed that there is a correlation between the X-ray luminosity corresponding to the hard-to-soft transition and the X-ray luminosity of the following soft state. This correlation holds over a luminosity range spanning by two orders of magnitude,  with no indication of a flux saturation or cut-off. We have also found that the transition luminosity correlates with the rate of increase in the X-ray luminosity during the rising phase of an outburst or flare, implying that the origin of the variation of the transition luminosity  is associated with non-stationary accretion in both transient sources and persistent sources. The correlation between the luminosity corresponding to the end of the soft-to-hard transition and the peak luminosity of the preceding soft state is found insignificant. The results suggest that the hysteresis effect of spectral state transitions is primarily driven by non-stationary accretion when the mass accretion rate increases rather than the mass accretion rate decreases. Our results also imply that  Galactic X-ray binaries can reach more luminous hard states during outbursts of higher luminosities and of similar rise time scales as those observed. Based on the correlations, we speculate that bright hard state beyond the Eddington luminosity will be observed in Galactic binaries in the next century. We also suggest that some ultra-luminous X-ray sources in nearby galaxies, which stay in the hard states during bright, short flares, harbor stellar-mass compact stars. 
\end{abstract}


\section{Introduction}
Observations of black hole and neutron star X-ray binaries show two main X-ray spectral states, namely the soft state and the hard state (see the recent review by Remillard \& McClintock 2006 and references therein). In the soft state, the X-ray spectrum is dominated by a thermal component with a weak steep power-law component. In the hard state, the X-ray spectrum is well described by a single power-law component with a high-energy cut-off (see the review by Done et al. 2007). For a long time it has been believed that a change in the instantaneous mass accretion rate around a threshold causes spectral state transitions,  therefore mass accretion rate was considered as the dominant parameter in determining spectral state transitions (e.g., Esin et al. 1997). 

Observations suggest that mass accretion rate is not the only parameter in determining spectral state transitions. The luminosity corresponding to the hard-to-soft (here after H-S) transition is found usually higher than that of the soft-to-hard (here after S-H) transition in transient sources \citep{miyamoto95,nowak95,mc03,gladstone07}, suggesting that the transitions between the two states do not occur at the same luminosity when transition directions are opposite. On the other hand, additional studies show evidence that the H-S transition luminosity is not fixed \citep{homan01, zd04}, which could vary by a factor of 10 \citep{yu04,yd07}. Many different parameters have been suggested to determine spectral state transitions, including the size of the corona \citep{homan01}, the history of the mass accretion rate  \citep{homan05}, the history of the location of the inner disk radius \citep{zd04}, two different flows in the accretion geometry which contribute to the non-thermal and thermal components, respectively \citep{smith02}, the mass in the accretion disk \citep{yu04, yu07}, and the changes in the Compton cooling and heating processes in the accretion flow involving evaporation and irradiation \citep{meyer04,liu05}, etc. These diverse suggestions likely reflect different manifestations of the same phenomena and  the same physics underneath. 

Usually during the rising phase of a bright outburst of a neutron star or black hole transient, spectral transition from the hard state to the soft state can be seen. A hard flare before the H-S transition occurs in neutron star and black hole transients \citep[see][and references therein]{bo84, br02, yu03}. The transition luminosity from the hard state to the soft state corresponds to the peak luminosity of the hard state throughout an outbursts because of hysteresis effect of spectral state transitions \citep{miyamoto95}. The transition does not occur at a constant luminosity nor at an arbitrary one. It has been found that the peak luminosity of the hard state correlates with the outburst peak luminosity in a number of soft X-ray transients Aql X-1, XTE J1550-564, and GX 339-4 and the persistent, transient-like neutron star low mass X-ray binary 4U 1705-44 \citep{yu04,yu07,yd07}. The general picture is that the brighter the initial hard state, the brighter the outburst will be, or vice versa. An application of the  correlation is that one can predict the outburst peak luminosity during the rising phase of an outburst when the H-S transition occurs \citep{yu04,yu07,yd07}.  

The origin of such a correlation is not well understood, but the correlation has important implications. First, the widely-accepted idea that the mass accretion rate in the accretion flow determines spectral state and state transition is not complete. In Aql X-1, \citet{yd07} showed that the correlation holds for outburst peak fluxes spanning by an order of magnitude, with the lowest transition flux comparable to that of the S-H transition. The correlation therefore holds for low luminosity flares. This suggests that spectral state transitions in bright outbursts or small flares could be consistently explained. Second, the correlation has shown no saturation towards high luminosities, which suggests that hard states brighter than the ones currently known are possible and could be observed in transient sources during brighter outbursts, leaving more luminosity room for the hard state to develop. Models aiming at explaining H-S transitions or the brightest hard state a source can reach therefore need to consider the correlation in the first place. 

Up to now there have been a number of transient sources seen to stay in the hard state throughout their outbursts (e.g. the last outbursts of XTE J1550-564, see \citep{be02}). This can be understood as that the mass accretion rate have not exceeded the threshold for state transitions predicted in the advection-dominated accretion flow (ADAF) model (e.g., Narayan \& Yi 1994,1995; Esin et al. 1997).  The correlation, on the other hand, suggests that a source can reach higher luminosities in the hard state than the theoretical threshold. The correlation and its intrinsic scatter then gives the peak luminosity of the thermal disk that is required for a transition to occur when a luminosity of the hard state is given; dimmer thermal disk could not develop and exist.   

In order to systematically study spectral state transitions to understand the origin of the luminosity correlation and if the correlation holds for not only transient sources but also persistent sources, we have performed a study of the spectral state transitions that can be seen in the long-term monitoring light curves of bright X-ray binaries from the All Sky Monitor (ASM) on board the Rossi X-ray Timing Explorer (RXTE) and the Burst Alert Telescope (BAT) on board the Swift in the 2--12 keV and the 15--50 keV energy ranges, respectively, in a period of three years. We searched for H-S transitions and S-H transitions in all the bright X-ray binaries and identified them based on hardness ratios between the BAT flux and the ASM flux. We confirmed the correlation between the luminosity corresponding to the H-S transition and the peak luminosity of the following soft state, and found the same correlation holds for both transient and persistent sources. We also show that the transition luminosity correlates with the rate of increase of the X-ray luminosity around the H-S transition, implying that the rate-of-increase of the mass accretion rate rather than the mass accretion rate itself drives the H-S transition during outbursts or flares primarily. This strongly suggests that most spectral state transitions observed in persistent and transient X-ray binaries should be explained with non-stationary accretion theories. 

\section{Observations and Data Analysis}
Public daily averaged X-ray light curves of bright X-ray binaries were obtained with the BAT (15$-$50 keV) and the ASM (2-12 keV). Combined together, they provide a monitoring of the X-ray sky on a daily basis, capable to be used to detect state transitions in bright X-ray binaries. 

We have searched for H-S transitions and S-H transitions in all the bright X-ray binaries. We took data from February 12, 2005 (MJD 53413) to February 8, 2008 (MJD 54504), the time when this study started. We used good BAT data with data\_flag (see Scaled Map Transient Analysis Synopsis of Swift/BAT\footnotemark{}) \footnotetext{http://heasarc.gsfc.nasa.gov/docs/swift/results/transients/Transient\_synopsis.html}) of 0 and dither\_flag of 0. The X-ray flux reported in this paper has been converted into units of Crab. We used 1 Crab = 75 count/s for the ASM and 1 Crab = 0.23 count/s/cm$^{2}$ for the BAT, estimated from the Crab light curves from the two instruments, respectively. We took advantage of the energy bands of the ASM and the BAT, which cover primarily for the thermal spectral component and the non-thermal spectral component, respectively. The hardness ratios between the ASM and the BAT fluxes then provide a comparison between the thermal and the non-thermal spectral components, which were proved very useful to determine spectral states. To increase detection sensitivity, we calculated two-day averaged results. We excluded BAT or ASM average rates with a significance smaller than 1$\sigma$. 

\subsection{Spectral states and measurements of transition fluxes}
We studied the ASM light curve, the BAT light curve and the hardness ratio for each source monitored by the RXTE/ASM and the Swift/BAT. We have identified spectral state transitions in bright neutron star and black hole binaries. The two-day averaged hardness ratios of these bright binaries show clearly two spectral states, namely the hard state and the soft state, respectively. In Figure 1, we plot the histograms of the distribution of the BAT/ASM ratios of 4U 1608-52 (upper left) and GX 339-4 (upper right), representing the neutron star systems and the black hole systems, respectively. All neutron star binaries and all black hole binaries (excluding GRS 1915+105) with spectral state transitions are shown on the lower left and lower right. A model composed of two Gaussian functions (on the scale of logarithm of the hardness ratio) can be used to fit the distributions. We determined the hardness ratio thresholds for the hard states and the soft states based on these histograms. The hard state hardness ratio threshold was taken as 1.0 for both neutron star and black hole systems, and the soft state hardness ratio thresholds were set to 0.2 and 0.1 for neutron stars and black holes, respectively.  These thresholds were marked as dashed and dotted lines in Figure 1. The thresholds 0.2 and 1.0 for neutron star systems correspond to the distribution peaks of the two spectral states. The broad, overlapping Gaussians for the two states suggest that reaching these hardness ratios would mean the sources have fully entered soft state or hard state for sure. For black hole systems, the soft state threshold was taken as 0.1, consistent with the soft state peak but smaller enough  than the hardness ratios corresponding to the very high or intermediate state (around 0.1-- 0.2) that occasionally seen during the outbursts of black hole transients. The hard state threshold was taken as 1.0, the same as those neutron star systems. When the hardness ratios are in the range above 1.0, black hole sources were definitely in hard states since the two Gaussians for the soft state and the hard state do not overlap in this hardness range. Special thresholds were chosen for the peculiar source GRS 1915+105 (see below and Figure 18), because its hardness ratio varied drastically on short time scales. 

When the H-S transition occurs, the hardness ratio shifts from above the hard state threshold to below the soft state threshold. We identified the hard X-ray flux corresponding to the start of each H-S transition from the BAT light curve and the peak flux in the soft X-ray corresponding to the following soft state from the ASM light curve according to the following scenarios. The transition flux from the hard state to the soft state was chosen as the BAT peak flux of the hard states around the H-S transition (within 6 days). This is based on previous studies of transient outbursts that the brightest hard state during the rising phase of an outburst corresponds to when the transition occurs \citep{yu03,yu04}. In case the flux variation is little,  the transition flux should correspond to that of the latest hard state before the transition occurs \citep{yd07}. The peak flux of the following soft state was chosen as the ASM peak flux of the soft state immediately after the H-S transition, but we exclude those isolated ASM peaks detected only in a single time bin to avoid false identifications. The fluxes we selected are therefore good approximations of the transition fluxes and peak fluxes.  

From Figure 2 to Figure 21, we show the long-term X-ray light curves of the bright Galactic X-ray binaries in which spectral state transitions were identified. For each H-S transition, we marked the start of the transition in the BAT light curve and the hardness ratio plot with a thin arrow, and the peak flux of the following soft state after the transition with a thick arrow, respectively. Whenever the S-H transition associated with the above H-S transition could be identified, we marked the start and the end of the S-H transition with thin and thick arrows, respectively, but with a dash-dotted tail to distinguish from those marks of the H-S transitions. We determined the transition flux from the soft state to the hard state by measuring the flux of the first hard state just after the hardness ratio changed from the soft state to the hard state threshold. Notice that the measured flux corresponds to the end of the S-H transition, which is different from those for the H-S transitions. If we assume that the S-H transition usually occurs during a luminosity decline, then the measured flux should be lower than the transition flux from soft state to hard state. 

The significance of the fluxes we identified was required to be above 3 $\sigma$. In Table 1, we describe the details of the state transitions identified in fifteen neutron star LMXBs and four black hole LMXBs. We also include the high mass X-ray binary Cyg X-3 because of its similarity to GRS 1915+105, although it does not show clear spectral states as those neutron star LMXBs and the black hole binaries GX 339-4 and GRO J1655-40. Other sources could not be added because of either a lack of simultaneous BAT/ASM data or a lack of identifications of state transitions. 

In summary, we identified H-S transitions and associated S-H transitions in fifteen neutron star LXMBs, four black hole LMXBs and a HMXB Cyg X-3. In order to compare them with the state transitions in the well-known black hole binary Cygnus X-1, we also estimated the transition flux and the peak flux of the following soft state during its H-S transition in 1996 based on CGRO/BATSE and RXTE/ASM monitoring observations \citep{zhang97}. 

\subsection{Measurement of the rate-of-change of the X-ray flux}
It has been suggested that an additional parameter is needed to account for state evolution based on studies of hysteresis \citep{homan01,smith02}. The unknown second parameter other than the mass accretion rate is the key for the understanding of the accretion geometry and the origin of spectral transitions. As suggested by the  correlation between the luminosity of the H-S transition and the peak luminosity of the following soft state, the second parameter could be the surrogate of the peak luminosity of an outburst or flare, such as the mass in the accretion disk before an outburst or flare occurs \citep{yu04,yu07}. In the simplest picture that the outburst rise time scale remains approximately constant in a single source, the rate-of-increase of the mass accretion rate (here after ${\rm d{\dot{M}}/{dt}}$), could be the surrogate of the peak luminosity of an outburst or flare. It is thus necessary to investigate the rate-of-change in the X-ray flux (here after ${\rm dL/dt}$) from the observations.

The idea that mass accretion rate determines spectral state is based on the assumption of stationary accretion under which the rate-of-change in the mass accretion rate has little effect. During transient outbursts this does not hold since the observed source flux usually increases by a factor of more than an order of magnitude in a period of a few days to a week. On the time scale of state transition, the increase of the mass accretion rate is no longer small compared to the mass accretion rate itself. 

We measured the rate-of-changes of the ASM and the BAT fluxes around the H-S transition, respectively, by calculating the ratio between the flux difference $\Delta{F}$ between any two adjacent time bins over the time interval of two days. In order to reduce the effects of large fluctuations from individual flux measurements, we performed numerical differentiation using 3-point, Lagrangian interpolation to measure the rate-of-change of the X-ray flux. For each H-S transition, we chose a time window to investigate the rate-of-change. The time window used for the BAT light curve starts when the BAT flux reached half of the BAT transition flux and ends when the BAT flux reached its maximum (transition flux). For  the ASM light curve, the time window starts at the same start time as the time window for the BAT analysis and ends at the ASM peak. We determined the maximal rate-of-changes in the ASM flux and the BAT flux in the above time windows, respectively, for each outburst or flare with H-S transition identified. 

The ${\rm dL/dt}$ in black hole binaries in the soft state or neutron star X-ray binaries in both soft state and hard state is an indicator of the ${\rm d{\dot{M}}/{dt}}$, because for the latter the gravitational energy of the matter accreted by neutron stars has to be released near or on the neutron star because of hard surface, and for the former the X-ray flux is dominated by the thermal emission from the accretion disk and is known to be a surrogate of the mass accretion rate in the disk flow. Close to the luminosity threshold for spectral transitions in the radiative inefficient flow model \citep{na94,na95}, which is the most popular model for black hole hard state, the radiation efficiency is proportional to the mass accretion rate and expected to be close to that of the standard disk flow, so the maximal rate-of-increase of flux approximately represent the maximal rate-of-increase of the mass accretion rate during the rising phase of the soft state or the hard state around the spectral state transition. This is supported by the following analysis, showing that black hole systems and neutron star systems fall on the same empirical correlation track (Fig.~24).

\section{Results}
\subsection{The hard-to-soft transition}
We have measured the hard X-ray flux corresponding to the H-S transition and the peak flux of the following soft state in about 20 bright sources. In Figure~22 we plot the observed transition flux and the observed peak fluxes of following soft states of these sources. It shows a strong positive correlation with Spearman correlation coefficient 0.91 and a chance possibility on the order of $10^{-30}$. To exclude the contribution due to diverse source distances and compact star masses, we re-scaled the observed fluxes to intrinsic fluxes to account for the effect from different source distances and compact star masses, using the source distances and compact star masses with uncertainties listed in Table 1. We found a strong positive correlation remains. 

Since intrinsic or Galactic absorption affects the observed soft X-ray flux below about 5 keV, we used only the ASM rate from the third channel (5-12 keV) to check if the flux correlation is primarily caused by this effect. We found a correlation of high significance with Spearman correlation coefficient of 0.85 and a chance possibility of $10^{-19}$. We concluded that intrinsic or Galactic hydrogen absorption and its potential variation does not play a role in causing the correlation between the transition fluxes and the peak fluxes. 

The energy spectra of Crab in the 2--10 keV band and the 15--50 keV band can be described by a power-law with photon indices of -2.07 and -2.12, and normalizations of 8.26 and 9.42, respectively \citep{kirsch05}. We can convert the ASM flux and the BAT flux in Crab units into luminosities assuming that the Galactic binaries have similar spectral shape as well as hydrogen absorption as the Crab. The approximation is justified since in the hard state just before a state transition, the energy spectrum is dominated by a power-law component with a power-law index in the range 1.8--2.0 and most of the binaries in our study have an $\rm N_{H}$ similar to that of the Crab ~$(4.0-4.5)\times{10}^{21}$. Figure 23 shows the relation between the luminosity corresponding to the H-S transition and the outburst peak luminosity. 
The Spearman coefficient is 0.85, with a chance possibility of $1.8\times{10}^{-20}$. We fit the data with a model of the form $\log {\rm L_{PS}}=A\log {\rm L_{tr,H}}$+B, where $\rm L_{PS}$ and $\rm L_{tr,H}$ represents the peak luminosity of the soft state and the H-S transition luminosity, respectively. We obtained A=1.06$\pm$0.06 and B=0.64$\pm$0.09, with an intrinsic scatter in $\log \rm L_{PS}$ of 0.18$\pm$0.002. The uncertainties in the estimates of source distances and masses, if known, are included during our model fits. But the data of Cyg X-3 was not used because we are not sure about whether it contains a black hole or a neutron star. These apply to the following fits as well. 

Figure 24 shows the correlation between the luminosities in Eddington units. We found a strong correlation between the transition luminosity and the peak luminosity of the following soft state. The Spearman correlation coefficient is 0.87, with the chance possibility being $7.4\times{10}^{-22}$. If we fit the data with a model of the same form as above, we obtained A=0.93$\pm$0.07 and B=0.42$\pm$0.14, with an intrinsic scatter in $\log \rm L_{PS}$ of 0.170$\pm$0.001. If only sources with estimates of distance uncertainties are considered, the Spearman correlation coefficient is 0.82, with the chance possibility being $9.6\times{10}^{-14}$. We obtained A=1.06$\pm$0.11 and B=0.65$\pm$0.19, with an intrinsic scatter in $\log \rm L_{PS}$ of 0.099$\pm$0.001. We also investigated if the correlation comes from several sources with the most spectral transitions detected. We excluded 4U 1728-34 and 4U 1636-53 in which more than ten transitions were identified. We found the Spearman correlation coefficient is 0.89, with a chance possibility of $1.5\times10^{-15}$. This suggests that the correlation between the luminosity of the H-S transition and the peak luminosity of the following soft state holds in X-ray binaries. It is worth noting that the H-S transition luminosity spans by about two orders of magnitude. As shown in the plot, the luminosity corresponding to the H-S transition in the 15--50 keV range is (0.13--8.0)\% Eddington luminosity. This corresponds to an X-ray luminosity (1--200 keV) of (0.4--25)\% assuming the source has a power-law spectrum with a photon index of 2 up to 200 keV when the H-S transition occurs. If the energy spectrum is a power-law spectrum with a cut-off at 30 keV, as sometimes seen just before a H-S transition, the X-ray luminosity in 1--200 keV would be 1/7 lower. This suggests that possible deviation from a single power-law spectrum for the hard state would not affect our luminosity estimates significantly. 

The correlation is found consistent with those determined previously in the outbursts of single sources Aql X-1, GX 339-4, and XTE J1550-564 \citep{yu04,yd07,yu07}. We took those fluxes measured with the RXTE ASM and HEXTE reported in these papers, converted them into Eddington units, and plotted those data in Figure 23 and 24 for the three sources (connected with a solid line for each source). The distance of XTE J1550-564 was taken as 5.3 kpc and the mass of the black hole was taken as 9.68 -- 11.58 ${M}_{\odot}$ \citep{or02}. The distance of Aql X-1 was taken as 5 kpc, as suggested by \citet{ru01}, and the mass was taken as 1.4 solar masses because the compact star is a neutron star. For GX 339-4 we used 5.6 kpc as its distance and a mass of 5.8 solar masses, which correspond to the lower limits. As shown in Figure 22, 23 and 24, the data measured in single sources with RXTE ASM and HEXTE follow the correlation track measured in this study very well. 

It is valuable to see where the H-S transition locates in these plots for the well-known black hole candidate Cygnus X-1, in which transitions are known to occur at similar luminosity levels \citep{zhang97,zd04}. We estimated the X-ray flux of the 1996 H-S transition and the peak flux of the following soft state using the ASM  and the BATSE fluxes. The ASM count rate of the soft state peak after the H-S transition was about 100 counts/s,  approximately 1.3 Crab, while the 20$-$50 keV flux estimated from BATSE corresponding to the start of the transition was about 1 crab \citep{zhang97}. The mass of the black hole in Cyg X-1 is about 10.1 $M_\sun$ and the distance is 2.1 kpc \citep{massey95}. Based on these parameters, we found the transition luminosity of Cyg X-1 locates at the low end of the transition luminosity range (see Fig. 24). The results suggest that Cyg X-1 has almost the lowest transition luminosity in the Galactic X-ray binaries. 

Based on the data shown in Fig. ~24, a comparison of the correlations in the neutron star systems and in the black hole systems can be made. GRO J1655-40 lies away from most of the data of the correlation track. This might be because of its unusual outburst profile or inclination angle, which does not allow consistent estimates of the fluxes. We took it as an outliner. We fit the data of the black hole systems including Cyg X-1, GRS 1915+105, GX 339-4 and XTE J1550-564, including results from previous studies. We obtained A=1.04$\pm$0.28 and B=0.48$\pm$0.45, with an intrinsic scatter in $\log \rm L_{PS}$ of 0.28$\pm$0.03. The Spearman correlation coefficient is 0.80. For the neutron star systems, we got A=1.03$\pm$0.05 and B=0.56$\pm$0.10, with an intrinsic scatter in $\log \rm L_{PS}$ of 0.092$\pm$0.001. The Spearman correlation coefficient is 0.88. The best fit models for the black hole systems and the neutron star systems are consistent with being the same. 

\subsection{The soft-to-hard transition }
In order to study the well-known hysteresis effect of spectral state transitions, we searched for the S-H transitions in those outbursts or low amplitude flares in which a H-S transition was  identified.  

Similar to the study of the H-S transitions, we have measured the 15--50 keV flux corresponding to the hard state immediately after the S-H transition. The advantage of selecting the hard state to measure transition luminosity is that we can compare the H-S transition flux and the S-H transition flux in the same energy range and in the same spectral state, which avoid highly uncertain bolometric corrections in the soft X-ray band for the soft state when the transition starts. The disadvantage is that we measure the hard state flux after the S-H transition has occurred. This is not the flux at which a source started a S-H transition, but the flux at which a source finished a S-H transition. If the S-H transition occurs during a luminosity decline, which is likely true in general, the measured flux should be significantly lower than the actual transition flux. 

We have converted the fluxes into luminosities in Eddington units and compared the transition luminosities between the S-H and the H-S transitions of the same outbursts or flares. We confirmed the hysteresis effect of state transitions that the S-H transition luminosity is generally lower than the H-S transition luminosity. This is shown in Figure 25. The 15--50 keV luminosity of the S-H transition is (0.2--2.0)\% Eddington luminosity, corresponding to an X-ray luminosity in 1--200 keV of about (0.6--6.0)\% Eddington luminosity. 

The luminosity of the S-H transition does not show a strong correlation with the preceding outburst peak flux. The corresponding Spearman correlation coefficient is 0.50, with a chance possibility of only 0.01, implying that there is low level positive correlation between the two but not significant. Examination of the data suggests that the correlation are largely contributed from 4U 1705-44, 4U 0614+091, and HETE J1900.1-2455. These data correspond to the upper and the lower luminosity ends of the date set, respectively. We also noticed that the uncertainties in our estimates of source distances and compact star masses should bring a scatter of source luminosities by a factor of 2 or so (e.g., if source distance estimate is uncertain by a factor of 1.4 or neutron star mass is 2.2 solar masses instead of 1.4 solar masses), which would contribute to a weak positive correlation in the plot.  Therefore it is unlikely that there is  a universal correlation between the S-H transition luminosity and the outburst peak luminosity among the sources from our analysis, but a correlation in single sources can not be ruled out (4U 1636-53, Homan 2007, private communication). It is worth noting that the S-H transition luminosity we defined is not the actual luminosity when the transition starts but the luminosity when the transition ends. 

\subsection{The hard-to-soft transition and the rate-of-increase of the X-ray luminosity}
We know that for outbursts or flares of similar rise time scales, outburst peak luminosity and the ${\rm dL/dt}$ correlates. For outbursts or X-ray flares of different rise times, outburst peak luminosity would not correlate with the ${\rm dL/dt}$ very well. Therefore, a study of the rate of increase in the X-ray luminosity vs. the H-S transition luminosity relation, as compared with the peak luminosity vs. the transition luminosity relation, would tell us which correlation is the primary correlation. In Figure 26, we plot the relation between the ${\rm dL/dt}$ and the H-S transition luminosity for the black hole and neutron star X-ray binaries as measured with the ASM and the BAT. We found a strong correlation between the two. For sources with estimates of distance uncertainties, We obtained Spearman correlation coefficients of 0.72 and 0.70, and chance possibilities of $2.4\times{10}^{-9}$ and $7.2\times{10}^{-9}$, for the ASM and the BAT measurements, respectively. We fit the data with a model of the form $\log \rm d{L}/{dt}=A\log {\rm L_{tr,H}}$+B, where $\rm d{L}/{dt}$ represents the rate-of-increase of the X-ray luminosity in the ASM or BAT. We obtained A=1.15$\pm$0.18, B=0.06$\pm$0.32 and an intrinsic scatter of 0.11$\pm$0.001 in $\log \rm d{L}/{dt}$ for the ASM and A=0.81$\pm$0.27, B=--1.16$\pm$0.45, and an intrinsic scatter of  0.10$\pm$0.002 for the BAT.

In Figure 27, the relation between the ${\rm dL/dt}$ and the peak luminosity of the following soft state is shown. Excluding sources without an uncertainty in the distance estimate, we found the Spearman correlation coefficients are 0.86 and 0.65, with chance possibilities of $1.9\times{10}^{-16}$ and $1.3\times{10}^{-7}$ , for the ASM and the BAT measurements, respectively. We fit the data with a model of the form $\log \rm d{L}/{dt}=A\log {\rm L_{PH}}$+B. We obtained A=1.18$\pm$0.14, B=--0.56$\pm$0.15, and an intrinsic scatter of $0.070\pm0.001$ in $\log \rm d{L}/{dt}$ for the ASM and A=0.67$\pm$0.25, B=--1.76$\pm$0.28 and an intrinsic scatter of $0.116\pm0.002$  for the BAT. 

We can not address whether it is the ${\rm dL/dt}$ or the outburst peak luminosity that primarily drives the correlations from the data set. However, the observations of the 2007 outburst in GX 339-4 might provide an evidence that the transition luminosity and the peak luminosity of the soft state have a weaker correlation than that between the transition luminosity and the rate-of-increase of the luminosity. During the 2007 outburst, the source reached a soft X-ray peak luminosity comparable to that of the 2002-2003 outburst as seen with the ASM, but its hard X-ray peak, as seen with the BAT, reached only half of the peak flux of the 2002-2003 outburst. The empirical relation would predict a peak luminosity of the outburst in the ASM to be half of the observed. This suggests that the outburst peak luminosity of the 2007 outburst deviated from the relation formed by the other three outbursts \citep{yu07} by a factor of 2.  The reason might be that the 2007 outburst has a significantly shorter rise time, which can be seen in the ASM light curve. 

\section{Conclusion and Discussion}
We have performed a systematic study of the state transitions in the brightest persistent and transient X-ray binaries observed with the RXTE/ASM and the Swift/BAT during a period of three years. We have found that the 15--50 keV luminosity corresponding to the H-S transition is positively correlated with the peak 2--12 keV luminosity of the following soft state and the rate-of-increase of the X-ray luminosity in a luminosity range spanning by two orders of magnitude (Figure 24 and 26). This does not only confirm the correlation previously found in single sources but also reveal that there is no clear cut in the state transitions between persistent sources and transient sources, suggesting that the observed large luminosity span of the H-S transition is caused by non-stationary accretion of  different scales, rather than the detailed mechanism of transient outburst and flare (e.g., inside-out or outside-in outbursts). This suggests that the additional parameter other than the mass accretion rate which determines spectral state transitions would relate to non-stationary accretion parameters, such as the ${\rm d{\dot M}/dt}$ or the initial condition such as the mass in the disk \citep{yu04,yu07}. On the other hand, both correlations show no luminosity saturation, suggesting that we have not observed the brightest hard state nor the brightest soft state which are permitted by physics. In other words, brighter hard states would be observed during brighter outbursts in the Galactic X-ray binaries. Such bright hard states and outbursts in stellar mass black holes may have already been observed in ultra-luminous X-ray sources (ULXs) in nearby galaxies. 

We also found that the luminosity corresponding to the end of the S-H transition does not show significant correlation with the peak luminosity of the preceding soft state in general. The transition flux as we defined is in general a few times lower than that of the H-S transition, but spans by more than one order of magnitude. This is larger than the factor of 4 difference estimated by \citet{maccarone03}, but uncertainties on the masses and distances and a narrower band used may account for the differences. 

\subsection{Luminosity of spectral state transitions and non-stationary accretion}
The correlation between the luminosity of the H-S transition and the ${\rm dL/dt}$ tells us that when there is little rate-of-increase of the X-ray luminosity, indicating little ${\rm d{\dot M}/dt}$ on the transition time scale, the transition luminosity of the H-S is the lowest. This is consistent with the low and constant transition luminosity seen in Cyg X-1 (Figure 24). Cyg X-1 stays at a rather constant flux level and shows similar transition luminosities \citep{zhang97, zd02, wilms06}. This can be understood as that Cyg X-1 has very little ${\rm d{\dot M}/dt}$ on the time scale of spectral state transitions. The implication of the correlation shown in Fig. 24 is that the difference in the H-S transition luminosity between the persistent sources and the transient sources lies in the difference in the ${\rm d{\dot M}/dt}$. Persistent sources tend to have lower ${\rm d{\dot M}/dt}$ in the hard state and therefore lower H-S transition luminosities; while the transient sources tend to have higher ${\rm d{\dot M}/dt}$ during outbursts. We speculate that the transition luminosity of the S-H transition is also affected by the rate-of-decrease of the mass accretion rate, as one could imagine that the faster a fading thermal disk disappears, the earlier the source enters the hard state, leading to the hard state to occur at a relatively higher luminosity. Whether the S-H transition luminosity is influenced by non-stationary accretion is worth further investigations. 

The tight correlation between the transition luminosity and the ${\rm dL/dt}$ suggests that there are two main parameters that determine the luminosity at which the spectral state transition occurs. One is the mass accretion rate, as generally suggested. The transition is determined by the mass accretion rate when there is little rate-of-increase of luminosity. The mass accretion rate sets the reference luminosity of the H-S transition corresponding to zero rate-of-change of the mass accretion rate. The other is the rate-of-increase in the mass accretion rate (note: whether there is strong effect of the rate-of-decrease of the mass accretion rate in the hard state is expected but not known yet, see Smith et al. 2007). In Figure 26, the slope between the transition luminosity and the ${\rm dL/dt}$ is determined. When the ${\rm dL/dt}$ is known, the range of the transition luminosity can be predicted. 

What causes sources in the hard states to reach higher luminosities than expected ? The answer is non-stationary accretion. This is set up by the initial condition at the beginning of an outburst or flare. The peak flux vs. waiting time relation found in GX 339-4 suggests that the hard X-ray peak luminosity is proportional to the mass stored in the disk (note: in GX 339-4 the mass stored in the disk during quiescence is approximately entirely accreted during the following outburst, see Yu et al. 2007).  The correlations between the transition luminosity and the peak luminosity of the following soft state or the ${\rm dL/dt}$ (Figure 24 and 26) indicate that the hard X-ray peak luminosity is positively correlated to the peak luminosity of an outburst or flare and the rate-of-increase of the X-ray luminosity. This links the rate of increase of the mass accretion rate to the mass stored in the disk before an outburst or flare occurs. How this link establishes is not clear. 

The results remind us that static accretion models are not good approximations of the accretion flow on the time scales when the variation of the mass accretion rate $\Delta \dot{M}$ is comparable to the mass accretion rate $\dot{M}$ itself.  The stationary condition was not met during most of the spectral state transitions during the rising phases of the outbursts or flares we studied. A complete state transition from the hard state to the soft state usually takes about a few days to a few weeks. As shown in Figure 26 and 27, the slope of the trend is $\sim$1, indicating that the daily increase of the X-ray luminosity during the rising phase of an outburst or flare is about 1/3 of the total luminosity itself. This indicates that during the rising phase of an outburst or flare, the mass accretion rate increases rapidly, and the variation of the mass accretion rate can not be treated as a small perturbation. This suggests that using static accretion solutions to interpret these spectral state transitions, as those models under the assumption of stationary accretion (for example, Esin et al. 1997; Meyer-Hofmeister et al. 2004;Liu et al. 2005), is questionable. The importance of time-dependent approaches is also supported by some other observations of spectral state transitions \citep{smith02, smith07}. 

\subsection{The brightest hard states of black hole and neutron star binaries} 
Because of the hysteresis effect of state transitions \citep{miyamoto95,nowak95, mc03,gladstone07}, the brightest hard state is usually reached during the rising phase of an outburst in transient X-ray binaries \citep{bo84,br01,yu03}. In our systematic study of the H-S transition in the bright X-ray binaries, the correlations do not show a flux saturation or cut-off at either end of the X-ray luminosity range, as shown in Figure 24 and 26, suggesting that the maximum luminosity that is permitted in the hard state has not been observed in the Galactic X-ray binaries. 

In the Galactic X-ray binaries we studied, the maximum transition luminosity in the 15--50 keV range is about 7--8\% $\rm L_{E}$ (assume Cyg X-3 is not a neutron star system). This corresponds to about $\sim$ 25\% Eddington luminosity in the energy range 1--200 keV assuming a power-law spectrum with photon index 1.8--2.0. If we consider that the hard X-ray peak flux of the 2002-2003 outburst of the black hole transient GX 339-4 is about twice of that of the 2006 outburst, the maximum H-S transition luminosity observed in the Galactic X-ray transients in the past decade is around 30\% (see also Zdziarski et al. 2004). 

How bright a black hole hard state can be ? What does the empirical relations tell us about the brightest hard state ? Based on Figure 24, hard states could reach higher peak luminosities in Eddington unit during outbursts or flares of higher peak luminosities. Therefore the chance to observe a much brighter hard state in these Galactic sources lies in whether there would be much brighter outbursts in future. 

We can roughly estimate the chance for us to observe a source in the hard state at the Eddington luminosity. In the past ten years or so we likely detected outbursts in the Galactic X-ray binaries on the order of 100 with peak luminosities higher than about 1\% ${\rm L}_{E}$. One out of these outbursts reaches 30\% $\rm L_{E}$ (e.g., the 2002--2003 outburst in GX 339-4). Assuming the distribution of outburst peak luminosity is of the form $N({\rm L_{p}})$=$A{\rm L_{p}}^{\alpha}$, where A is the normalization and $\alpha$ is the index, the chance for us to detect a hard state at Eddington luminosity would be ($1.0/0.3)^\alpha$ times the chance for us to see an outburst of 30\% ${\rm L}_{E}$ -- which is one in ten years. If $\alpha$ is close to -1, then we would observe an outburst with the hard state reaching the ${\rm L}_{E}$ in about 30 years' time. If $\alpha$ is close  to -1.5, then we would observe an outburst with the hard state reaching the ${\rm L}_{E}$ in about 60 years (see Grimm et al. 2002 for Galactic X-ray binary luminosity function). Similar estimate would apply to galaxies similar to our own. If $\alpha$ is in the range -1 -- -1.5, we would observed an X-ray binary in the hard state at the Eddington luminosity in about 3--6 galaxies during a period of ten years. This would account for some of the ultra-luminous X-ray sources (ULXs) seen in bright hard states in nearby galaxies. We found that the outburst sample in our current study is not enough for a determination of the distribution of the outburst peak luminosity. Future statistical studies of transient outbursts would determine the distribution and give a certain, quantitative answer. 

Based on the empirical relations, the outbursts or flares during which a much brighter hard state can be seen are those having a larger rate-of-increase of the luminosity than what have been seen in these  Galactic sources. To have a larger rate-of-increase of the luminosity, an outburst or flare should either reach higher peak luminosity within a similar rise time as the outbursts seen in the Galactic transients, or reach similar peak luminosity within a shorter rise time. Specific prediction about the interesting source GX 339-4 can be made. Based on the empirical relation between the peak flux of the hard state and the outburst waiting time \citep{yu07}, GX 339-4 will reach the Eddington luminosity in the hard state during a future outburst after being inactive for ten years. It worth noting that the peak flux vs. waiting time relation reported in \citet{yu07} was based on BATSE monitoring observations. Source activities below about 0.1 Crab did not obviously affect the empirical relation. Therefore if GX 339-4 stays inactive (below 0.1 Crab) for more than 10 years, the peak flux of the hard state in the next outburst would reach beyond  the Eddington luminosity.

Have super-Eddington outbursts with similar rise or decay time scales as those of the Galactic transients been observed in sources in nearby galaxies ? The answer is yes. We have performed a study of the bright outbursts of the Galactic X-ray binaries seen with the RXTE/ASM and found that the shortest e-folding rise time scale in each outburst is on average about 2 days. There is evidence that the shortest characteristic rise or decay timescales during the X-ray flares of some ULXs are comparable. One of the best examples is seen in NGC 1365 X-1, which declined in luminosity with an e-folding time scale of about 3 days during a series Chandra/ACIS snapshots \citep{soria07}. Another example is a supersoft source in M 101, which showed an outburst to the soft state with an e-folding time scale around 1--2 days \citep{kong04}.  Yet another example is a ULX in M82: a few X-ray flares were seen emerging from a 62-day flux modulation \citep{kaaret06}. We have investigated the RXTE/PCA light curve and found that the e-folding rise and decay timescales being about 2$-$3 days. These bright hard states observed during bright, short flares are expected from stellar mass black holes and neutron stars based on the correlations between the H-S transition luminosity and the outburst peak luminosity or the ${\rm dL/dt}$ in stellar-mass black hole or neutron star X-ray binaries. Our study provides an additional evidence that some of the ULXs contain stellar mass compact stars. 

\subsection{Outbursts entirely in the hard state }
A number of hard state outbursts have been observed in X-ray transients, such as XTE J1550-564, XTE J1118-480, IGR J17497-281, SWIFT J1753.5-0127, and Aql X-1 \citep{be02,br04,ro06,ro07,ra07}. In the framework that mass accretion rate determines spectral state transitions, these hard state outbursts can only be explained as that the mass accretion rate threshold is not reached. Then brighter outbursts with the hard state exceeding the threshold, roughly of 1-2\% $\rm L_{E}$ in 1--200 keV corresponding to the transition luminosity of Cyg X-1 (see Fig.~24), can not be explained. 

The correlations shown in Fig.~24 and Fig.~26 provide an explanation of bright hard state outburst. In the plot of Fig.~26, the data defines a band of the transition regime (with a half width as the intrinsic scatter) in the transition luminosity of the hard state vs. the rate-of-increase of the luminosity relation. Two theoretical possibilities therefore can be inferred for when the state transitions do not occur. One is that these sources stay in the lower-right regime below the transition regime and have had lower rate-of-increase of the luminosity in the recent past in relative to their peak luminosities of the hard states. Combined with the correlation shown in Fig.~24, the occurrence of hard state outburst is because of low ${\rm dL/dt}$ in the recent past which does not permit the source to reach a soft state of which the luminosity should follow the relation defined in Fig.~24, i.e., soft state of lower luminosity can not exist. The other possibility is that the sources stay in the upper-left regime in Fig.~26 and have lower peak luminosity of the hard state in relative to the rate-of-increase of the luminosity, suggesting that the hard state outbursts are due to lower peak luminosities in relative to the ${\rm dL/dt}$. This indicates a potentially existing, but special regime of the hard state under violent conditions of the accretion flow, which may account for not only bright hard state in Galactic binaries but also those seen in ULXs.  Combined with the correlation in Fig.~24, the occurrence of hard state outburst is because the source has a low luminosity in relative to the ${\rm dL/dt}$. Such hard states can sustain for some time but not forever since otherwise this would lead to an infinite luminosity. Therefore thermal disk with a luminosity higher than the value predicted in the relation in Fig.~24 can not develop. Whether this case is related to the occurrence of very high state or intermediate state in black hole systems is unclear, but clearly deserves further studies. 

Two hard state outbursts around MJD 53926 and 54618 (outside the time range of our analysis) in 4U 1608-52 can be seen in the ASM and BAT light curves. We can determine the peak flux of the hard state and the maximum rate-of-increase of the fluxes since the sources reached half of their peak fluxes for the two outbursts. The maximum rate-of-increase of the X-ray flux as seen in the ASM and BAT during these hard state outbursts is statistically lower by a factor of 4 compared with the outbursts during which the H-S transition occurs around MJD 54270 and 54400, but comparable to that occurred around MJD 53650. This seems to suggest that our approach with maximum rate-of-increase of the X-ray flux is too simple to reveal the nature of hard state outburst. 

\section{Summary}
We have studied spectral state transitions in the brightest persistent and transient Galactic X-ray binaries seen with the X-ray monitoring observations of the RXTE/ASM and the Swift/BAT. We have confirmed that the luminosity of the H-S transition is correlated with the peak luminosity of the following soft state, and found that the correlation holds for both persistent sources and transient sources in a luminosity range spanning by two orders of magnitude. We have also found the rate-of-increase of the luminosity is correlated with the transition luminosity or the peak luminosity of the following soft state. The results imply that state transitions occur in a large range of mass accretion rates and the majority of the H-S transitions observed are strongly influenced by non-stationary accretion. The main results can be summarized as follows:
\begin{itemize}
\item Both correlations do not show high luminosity saturation or low luminosity cut-off, suggesting that we have not observed the brightest hard state nor the dimmest soft state in the Galactic black hole or neutron star soft X-ray binaries. An outburst reaching the Eddington luminosity in the hard state would be observed in GX 339-4 if the source stays below $\sim$ 0.1 Crab for a period of about ten years.  
\item The correlations suggest that brighter hard state could be reached during the rising phase of a brighter outburst with similar rise time or similarly bright outburst with shorter rise time in stellar-mass black hole or neutron star transients. Several ULXs showing short-duration hard flares therefore likely harbor stellar-mass compact stars. 
\item The two correlations are dependent. The correlation between the rate-of-increase of X-ray luminosity and the H-S transition luminosity could introduce the correlation between the peak luminosity of the following soft state and the transition luminosity if the rise time scales are similar among outbursts or flares in single sources and across sources. 
\item The luminosity of the H-S transition is shown to correlate with the rate-of-increase of the luminosity, suggesting that it is non-stationary accretion, characterized by the rate-of-increase of the mass accretion rate, determines the variation of the transition luminosity. Combined with the results obtained from GX 339-4 \citep{yu07}, the rate-of-increase of the mass accretion rate is nearly proportional to the mass in the accretion disk involved in an outburst or flare. Cygnus X-1 is at the low luminosity end in the correlation tracks, consistent with previous suggestions that its mass accretion rate is in a narrow range which leads to low rate-of-increase of the mass accretion rate. 
\end{itemize}

The schematic picture concerning the allowed regimes of the luminosities of the H-S transition and the hard state is shown in Fig.~28. The light curves of two assumed outbursts of the same source are shown.  The solid curve represents the X-ray light curve of a brighter outburst while the dashed line represents that of a weaker one. The allowed regimes for the hard state (gray) or the soft state (white) are shown based on our results. The maximum luminosity permitted in either the hard state or the soft state is yet unknown. The popular picture based on the idea that the mass accretion rate determines spectral state predicts that the H-S and the S-H transitions occur at a constant luminosity $\rm L_{0}$, below which a source stays in the hard state. An additional hard state regime is on the rising phase of an outburst or flare in X-ray binaries based on the empirical correlations. The schematic picture shows that much brighter hard state can be reached during bright, short outbursts in stellar-mass black hole and neutron star binaries in our Galaxy.  We infer that Galactic binaries may turn into ULXs during shorter, brighter  outbursts. The picture also suggests that hysteresis effect of state transitions is mainly caused by the H-S transition strongly influenced  by non-stationary accretion characterized by the rate-of-increase of the mass accretion rate, of which the initial condition may be described by the mass in the accretion disk.

\acknowledgments
We would like to thank the RXTE and the Swift Guest Observer Facilities at NASA Goddard Space Flight Center for providing the RXTE/ASM products and the Swift/BAT transient monitoring results. We thank the anonymous referee for useful comments and suggestions. WY would also like to thank Roberto Soria of University College London and Albert Kong of National Tsing Hua University for sharing their studies on ultra-luminous X-ray sources, D. M. Smith of University of California at San Diego for sharing his work before publication, and Thomas Maccarone of University of Southampton and Diego Altamirano of University of Amsterdam for comments and careful reading of the manuscript. WY appreciate useful discussions with Chris Done, Robert Fender, Tomaso Belloni, Jean Swank, Ron Remillard, John Tomsick, Joern Wilms, and Mike Nowak. This work was supported in part by the National Natural Science Foundation of China (10773023, 10833002), the One Hundred Talents project of the Chinese Academy of Sciences, the Shanghai Pujiang Program (08PJ14111), the National Basic Research Program of China (973 project 2009CB824800), and the starting funds at the Shanghai Astronomical Observatory. The study has made use of data obtained through the High Energy Astrophysics Science Archive Research Center Online Service, provided by the NASA/Goddard Space Flight Center.

\newpage

\begin{figure*}
\epsscale{0.6}
\plotone{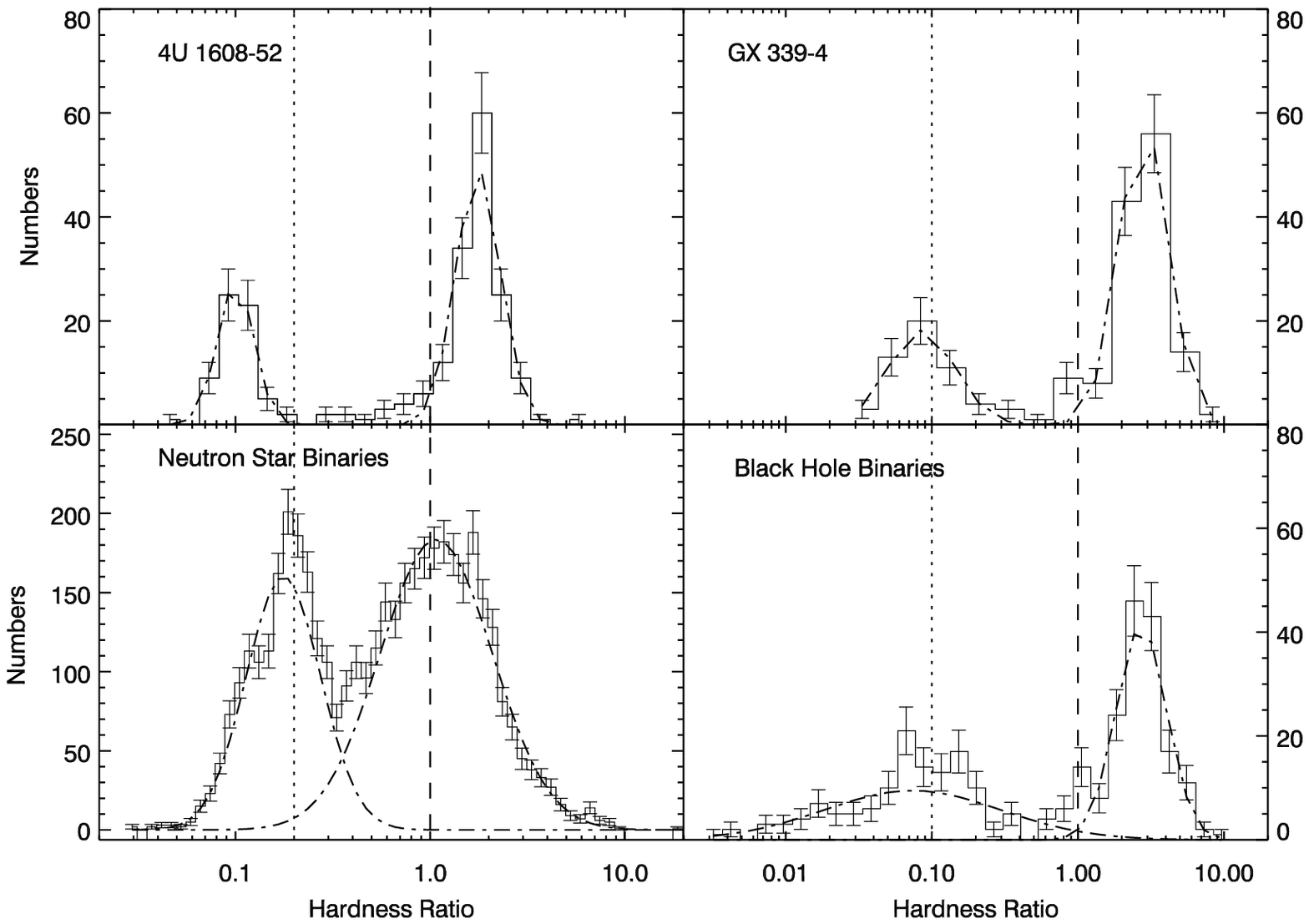}
\caption{The distribution of the BAT/ASM hardness ratio on time scale two days for the bright neutron star LMXBs and black hole binaries in which spectral state transitions were detected. 4U 1608-52 and GX 339-4 are the best examples showing distinct spectral states for neutron stars and black holes, respectively. Data of GRS 1915+105 were not included in the statistics of black hole binaries. The hardness thresholds for the soft states and the hard states which were used in our analysis are marked as dotted-lines and dashed-lines, respectively. }
\end{figure*}
\clearpage

\begin{figure*}
\epsscale{0.6}
\plotone{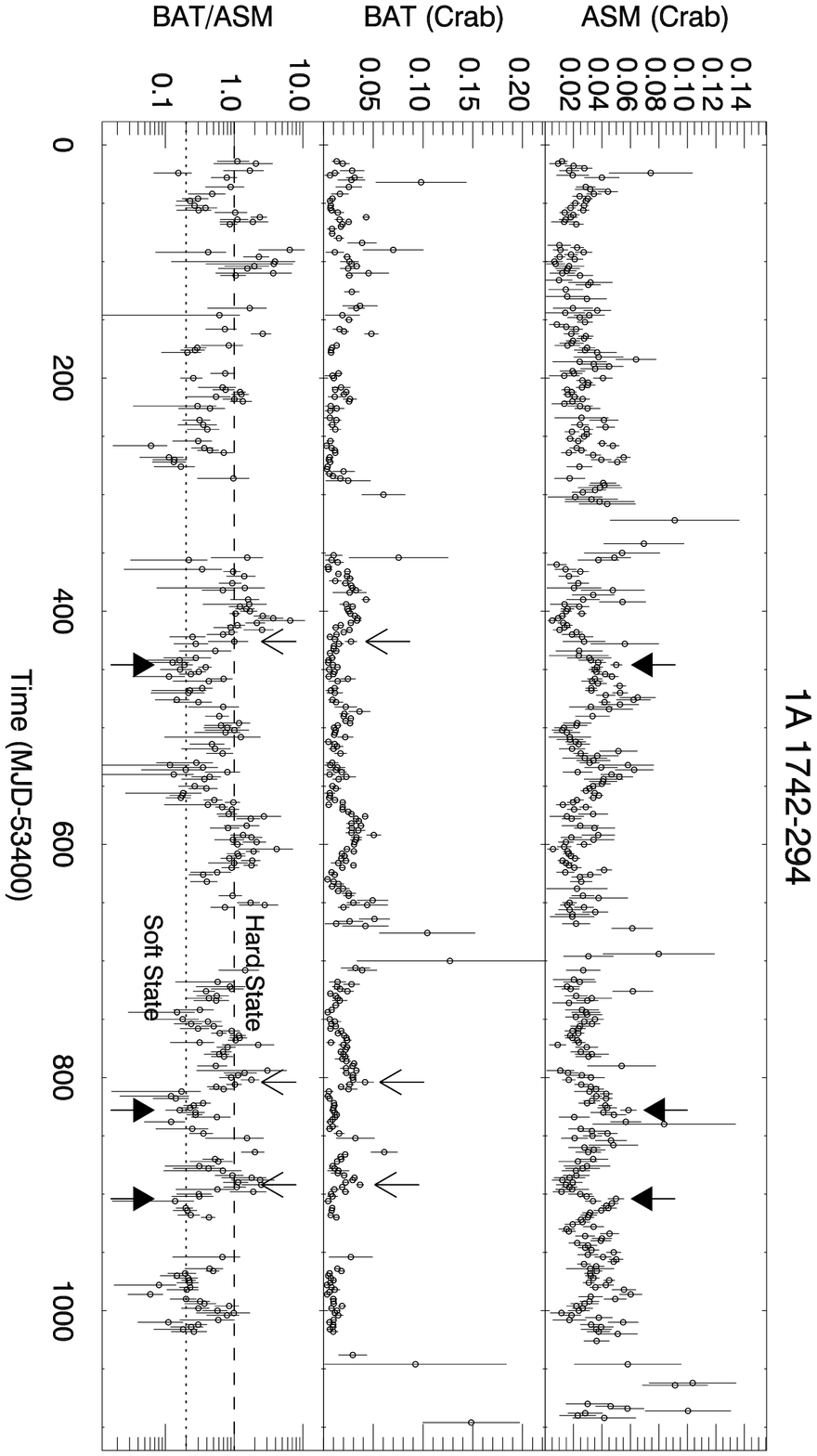}
\caption{X-ray monitoring observations of 1A 1742-294 in 2--12 keV with the ASM and 15--50 keV with the BAT.}
\end{figure*}
\clearpage

\begin{figure*}
\epsscale{0.6}
\plotone{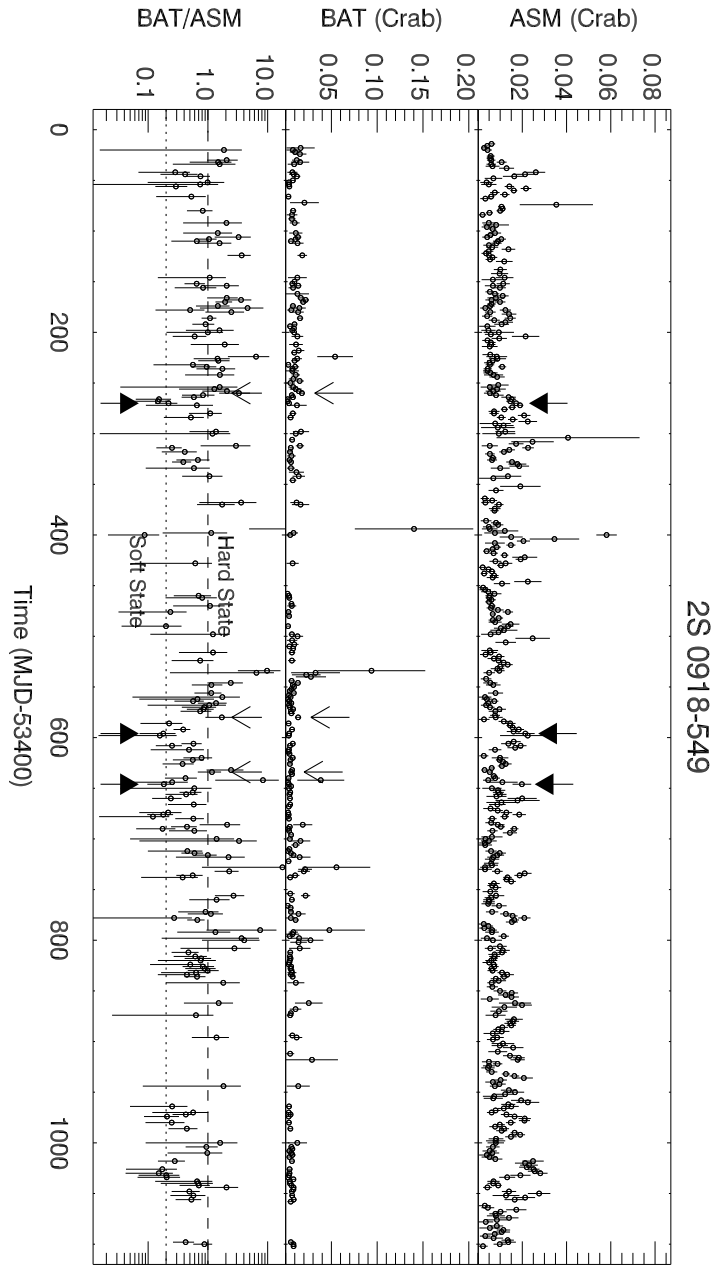}
\caption{X-ray monitoring observations of 2S 0918-549 in 2--12 keV with the ASM and 15--50 keV with the BAT.}
\end{figure*}
\clearpage

\begin{figure*}
\epsscale{0.6}
\plotone{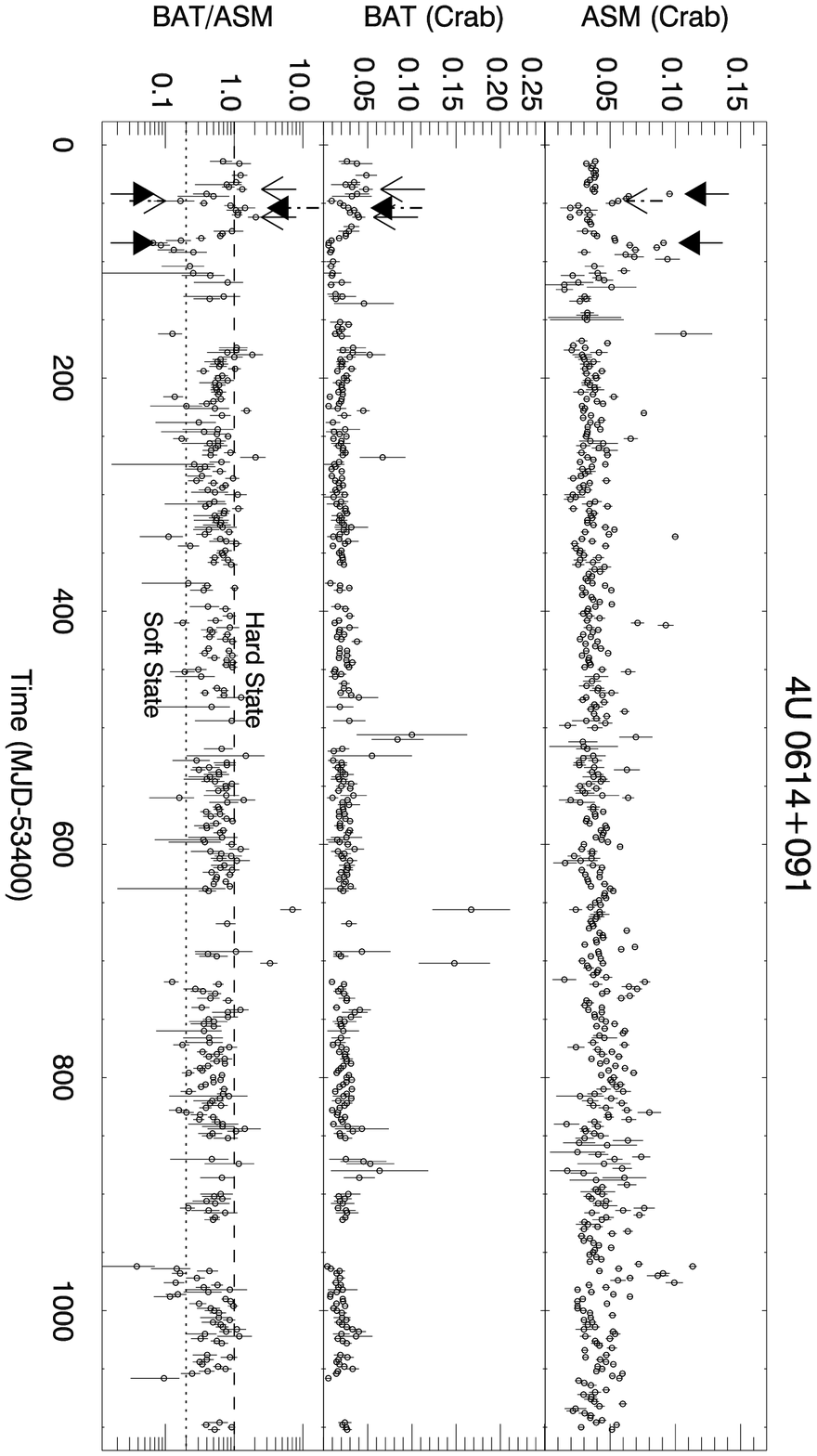}
\caption{X-ray monitoring observations of 4U 0614+091 in 2--12 keV with the ASM and 15--50 keV with the BAT.}
\end{figure*}
\clearpage

\begin{figure*}
\epsscale{0.6}
\plotone{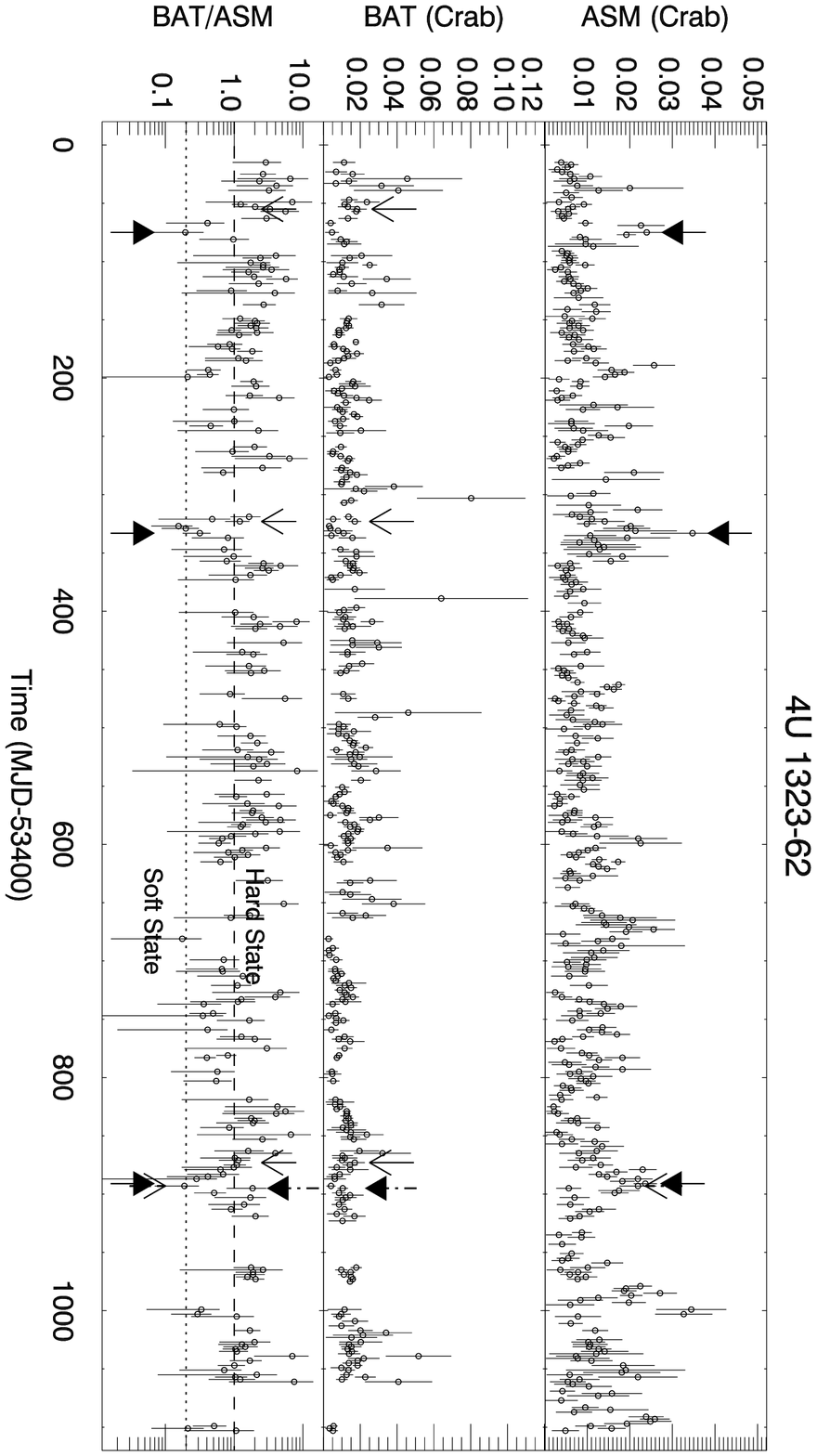}
\caption{X-ray monitoring observations of 4U 1323-62 in 2--12 keV with the ASM and 15--50 keV with the BAT. }
\end{figure*}
\clearpage

\begin{figure*}
\epsscale{0.6}
\plotone{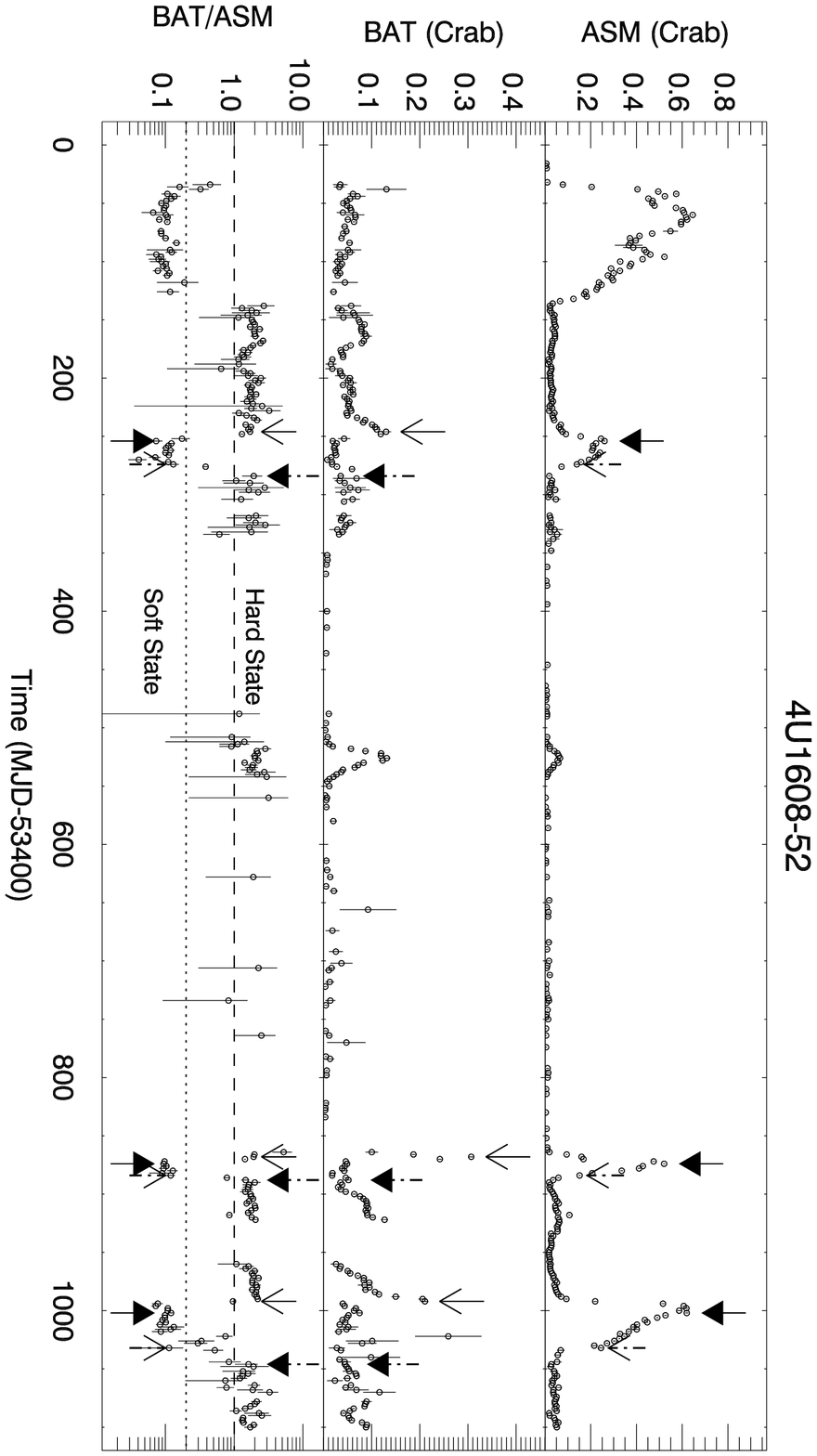}
\caption{X-ray monitoring observations of 4U 1608-52 in 2--12 keV with the ASM and 15--50 keV with the BAT. }
\end{figure*}
\clearpage

\begin{figure*}
\epsscale{0.6}
\plotone{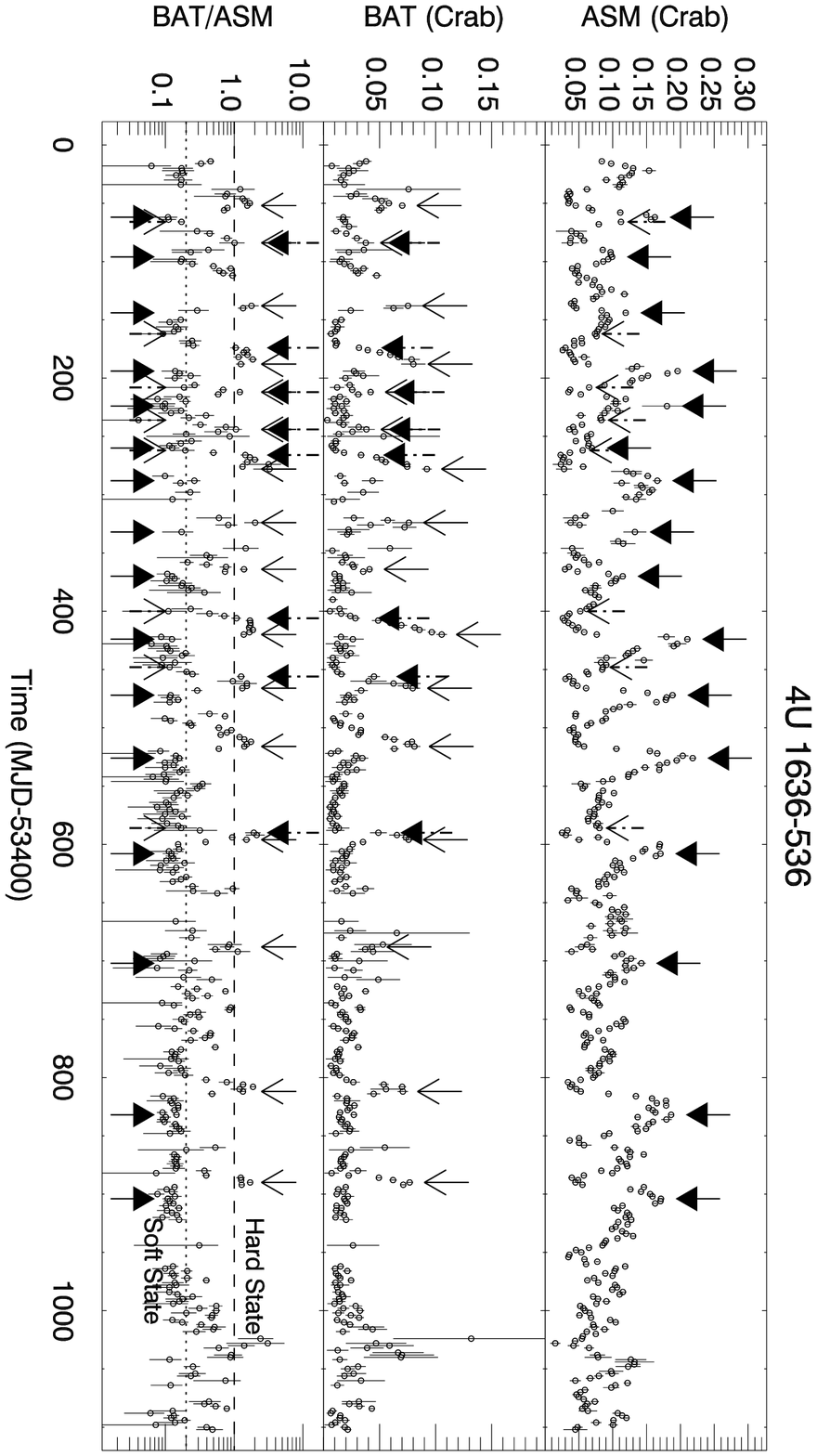}
\caption{X-ray monitoring observations of 4U 1636-53 in 2--12 keV with the ASM and 15--50 keV with the BAT.}
\end{figure*}
\clearpage

\begin{figure*}
\epsscale{0.6}
\plotone{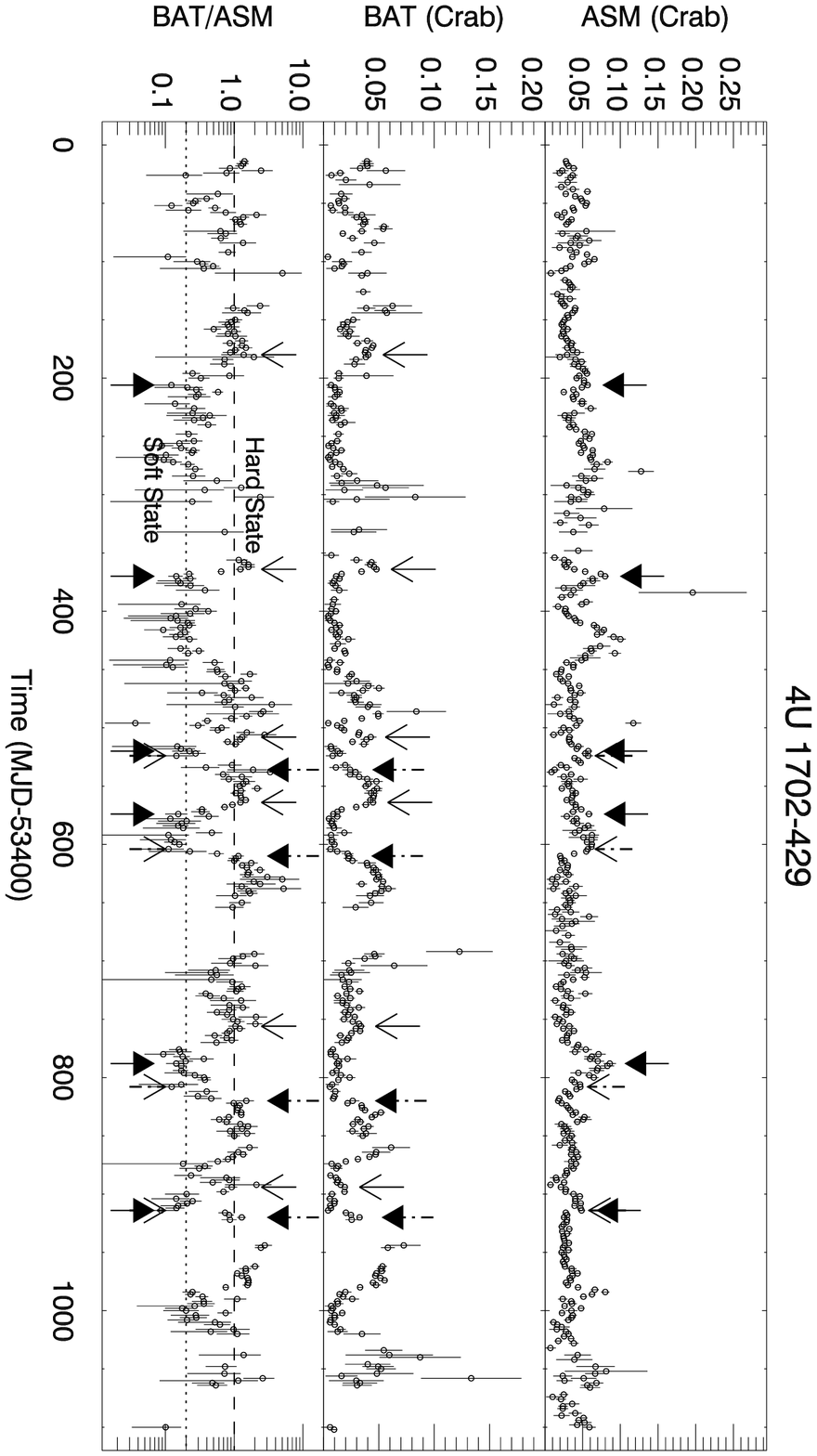}
\caption{X-ray monitoring observations of 4U 1702-429 in 2--12 keV with the ASM and 15--50 keV with the BAT. }
\end{figure*}
\clearpage

\begin{figure*}
\epsscale{0.6}
\plotone{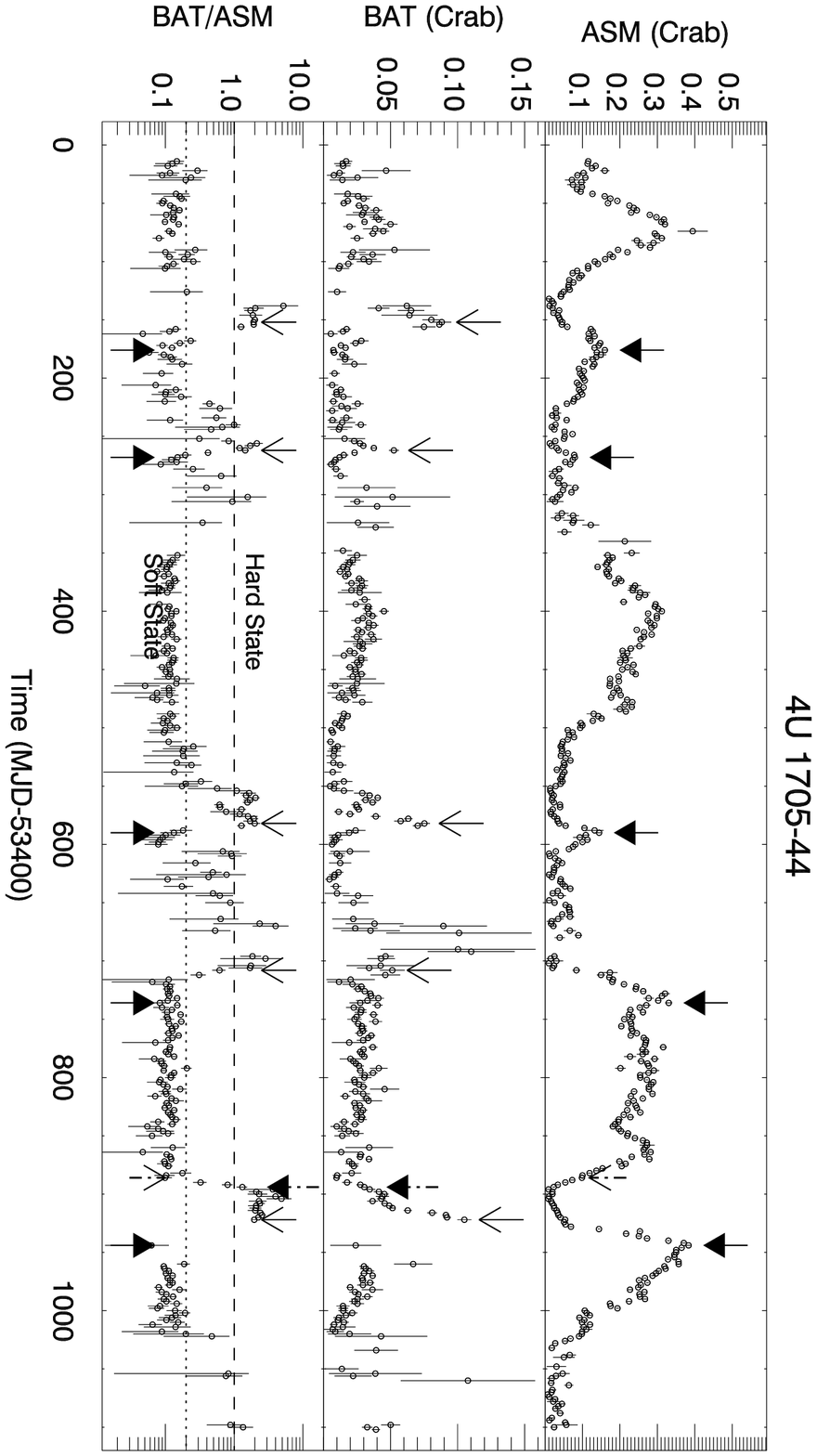}
\caption{X-ray monitoring observations of 4U 1705-44 in 2--12 keV with the ASM and 15--50 keV with the BAT. }
\end{figure*}
\clearpage

\begin{figure*}
\epsscale{0.6}
\plotone{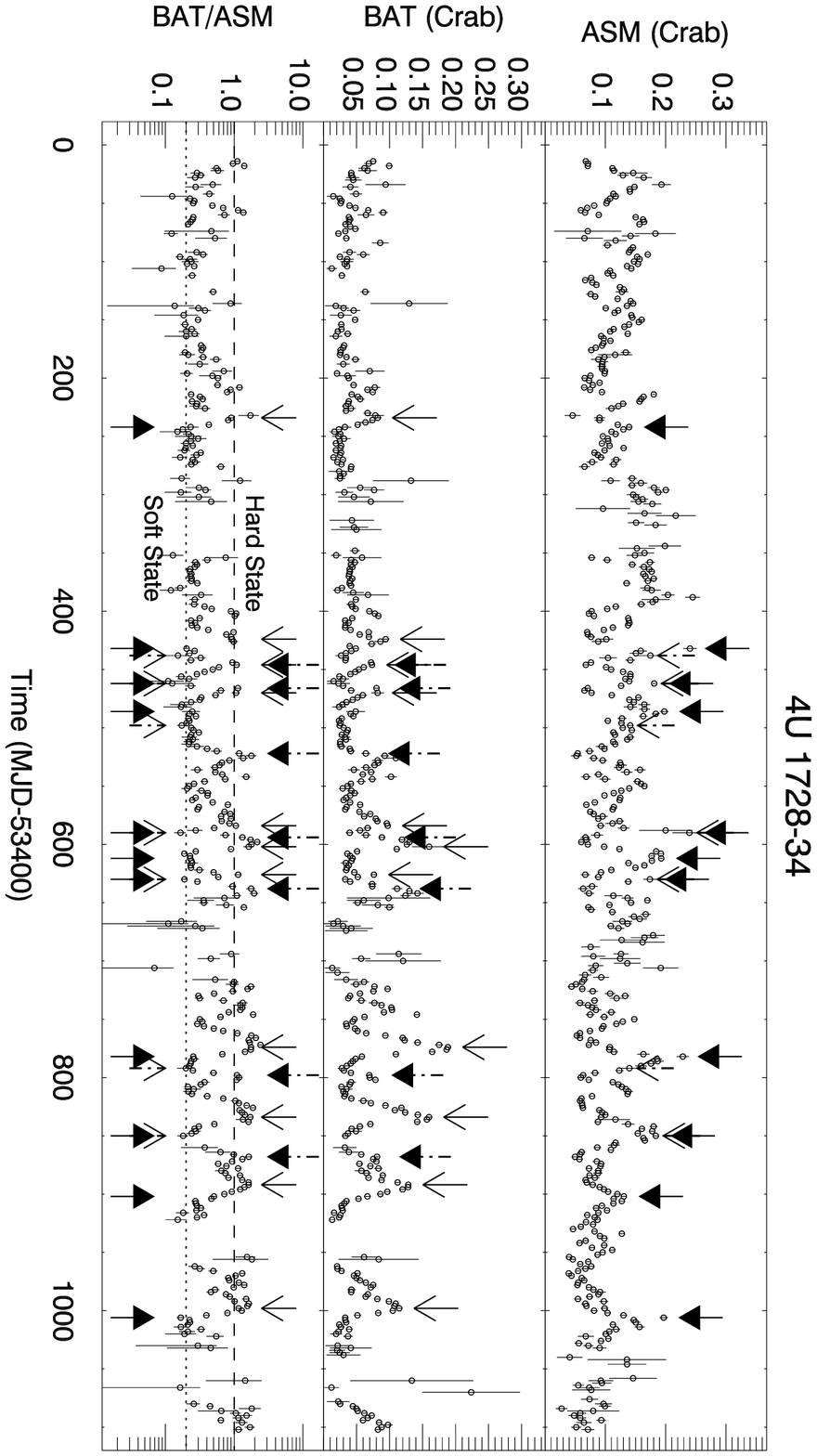}
\caption{ X-ray monitoring observations of 4U 1728-34 in 2--12 keV with the ASM and 15--50 keV with the BAT. }
\end{figure*}
\clearpage

\begin{figure*}
\epsscale{0.6}
\plotone{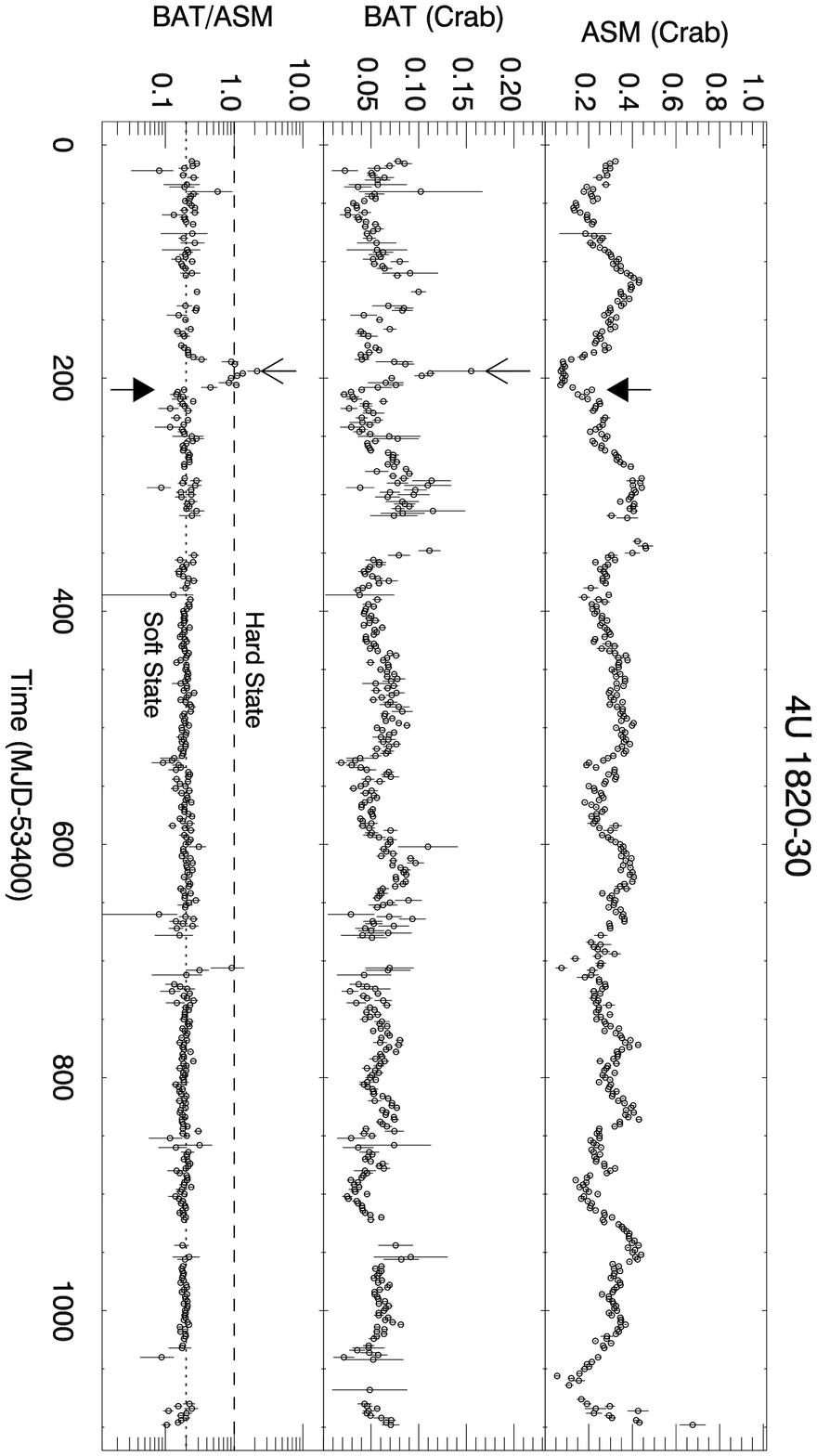}
\caption{X-ray monitoring observations of 4U 1820-30 in 2--12 keV with the ASM and 15--50 keV with the BAT.  }
\end{figure*}
\clearpage

\begin{figure*}
\epsscale{0.6}
\plotone{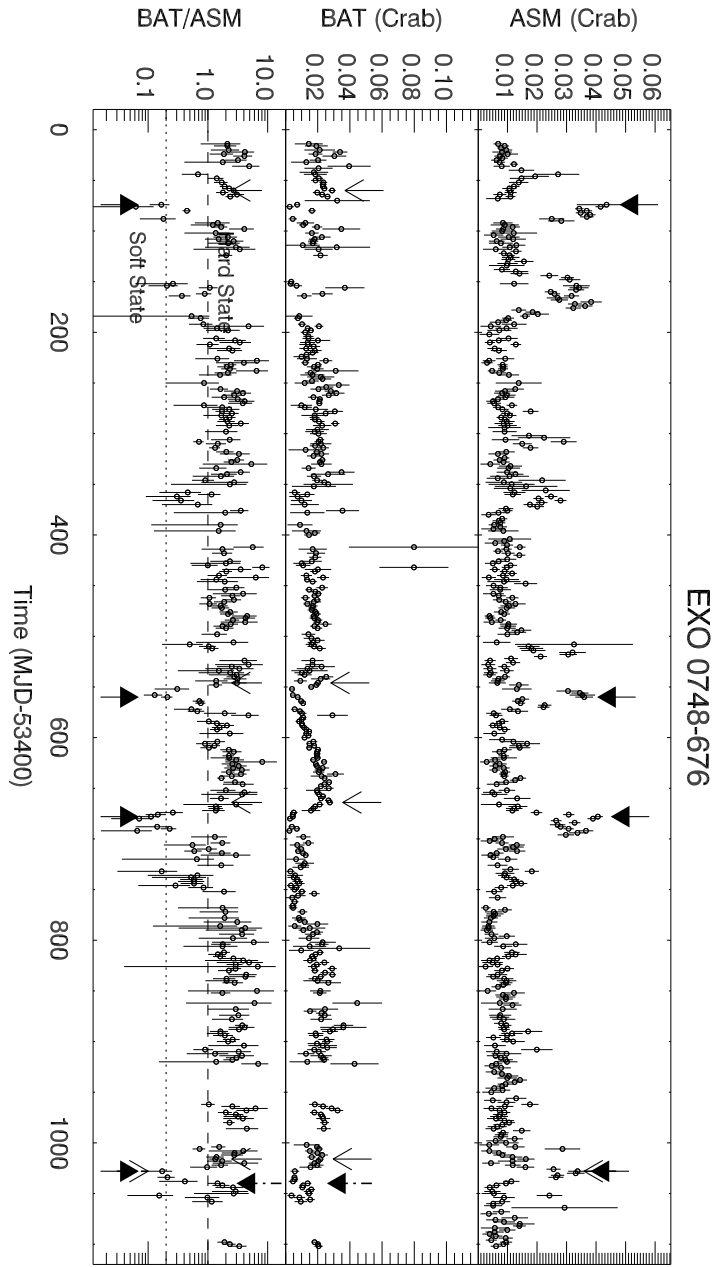}
\caption{X-ray monitoring observations of EXO 0748-676 in 2--12 keV with the ASM and 15--50 keV with the BAT. }
\end{figure*}
\clearpage

\begin{figure*}
\epsscale{0.6}
\plotone{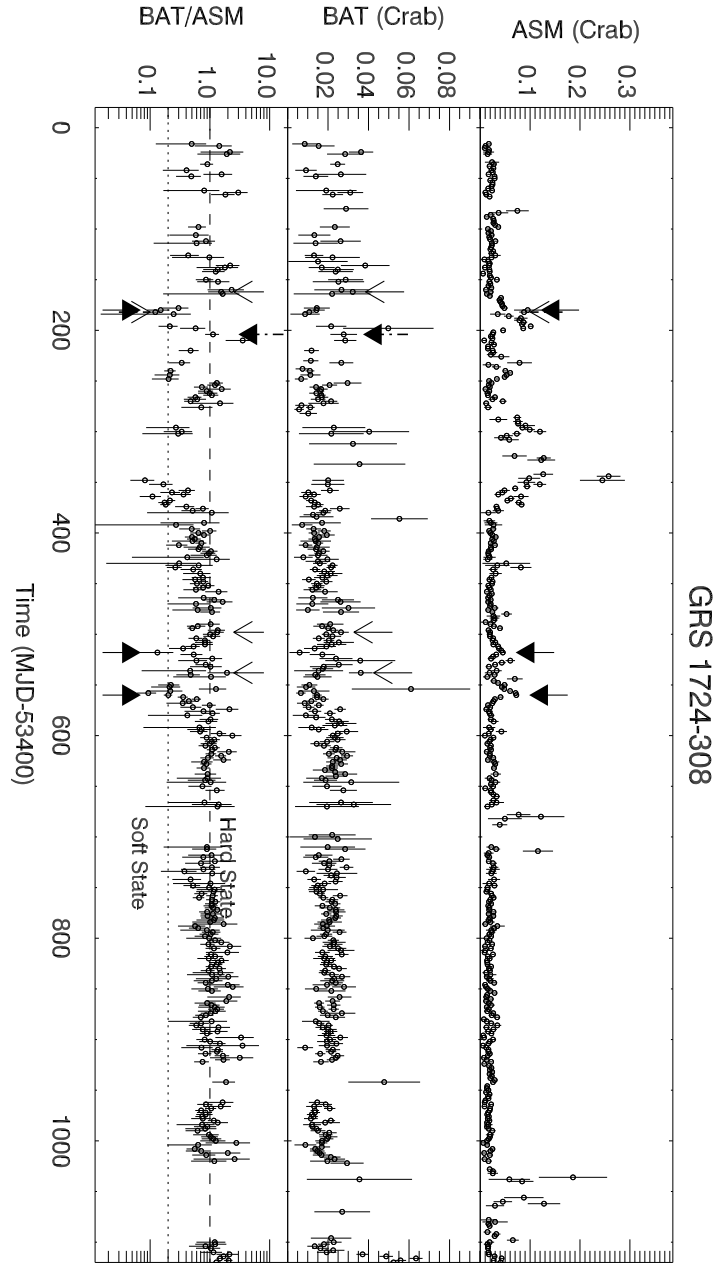}
\caption{X-ray monitoring observations of GRS 1724-308 in 2--12 keV with the ASM and 15--50 keV with the BAT. }
\end{figure*}
\clearpage

\begin{figure*}
\epsscale{0.6}
\plotone{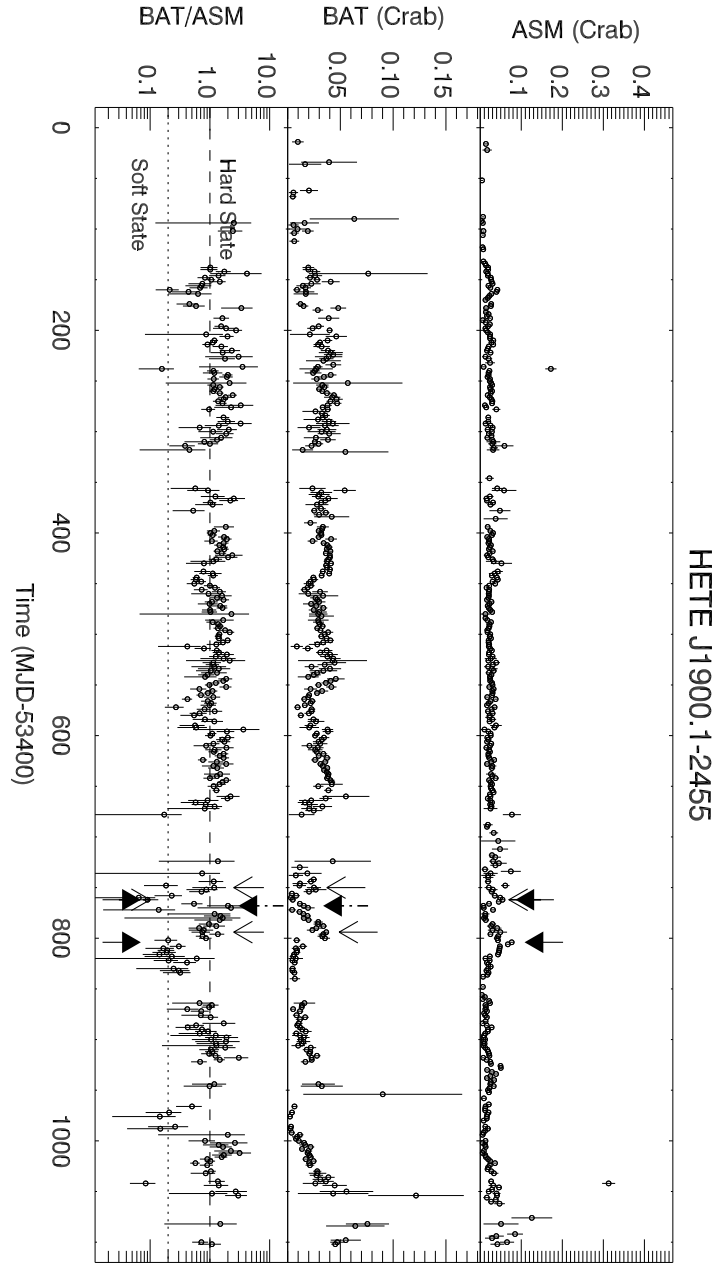}
\caption{X-ray monitoring observations of HETE J1900.1-2455 in 2--12 keV with the ASM and 15--50 keV with the BAT. }
\end{figure*}
\clearpage

\begin{figure*}
\epsscale{0.6}
\plotone{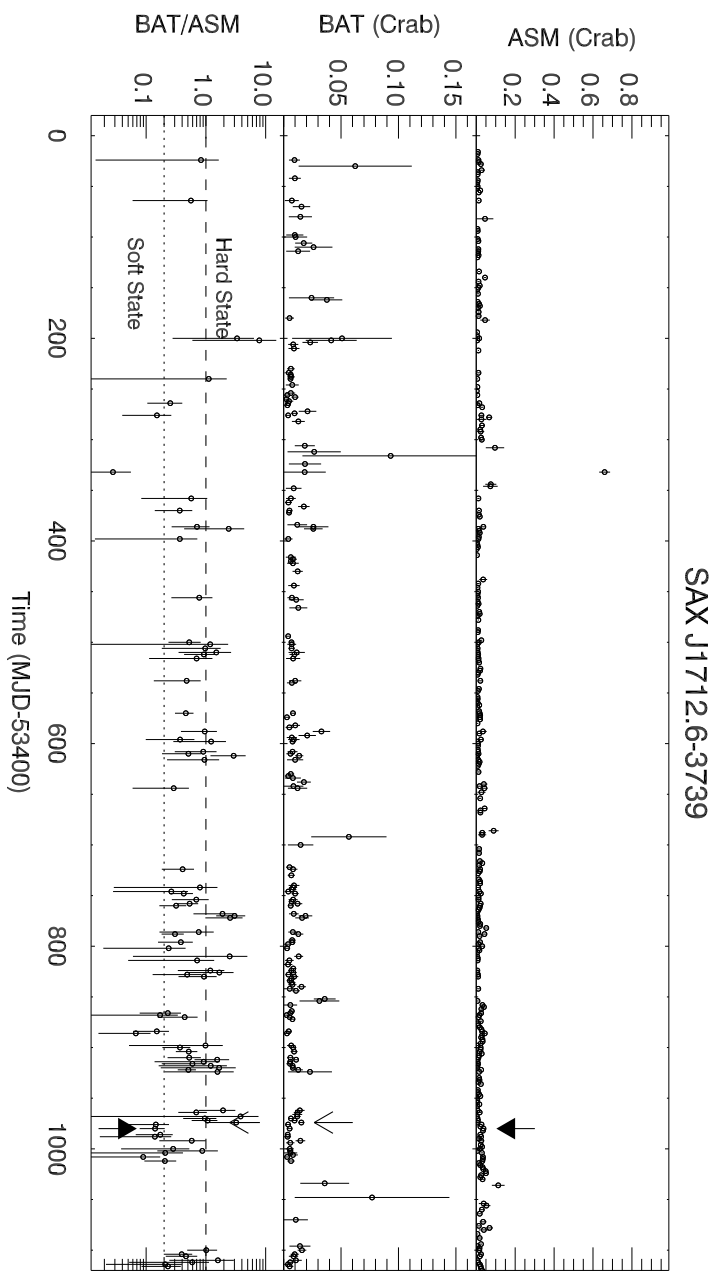}
\caption{X-ray monitoring observations of SAX J1712.6-3739 in 2--12 keV with the ASM and 15--50 keV with the BAT. }
\end{figure*}

\clearpage
\begin{figure*}
\epsscale{0.6}
\plotone{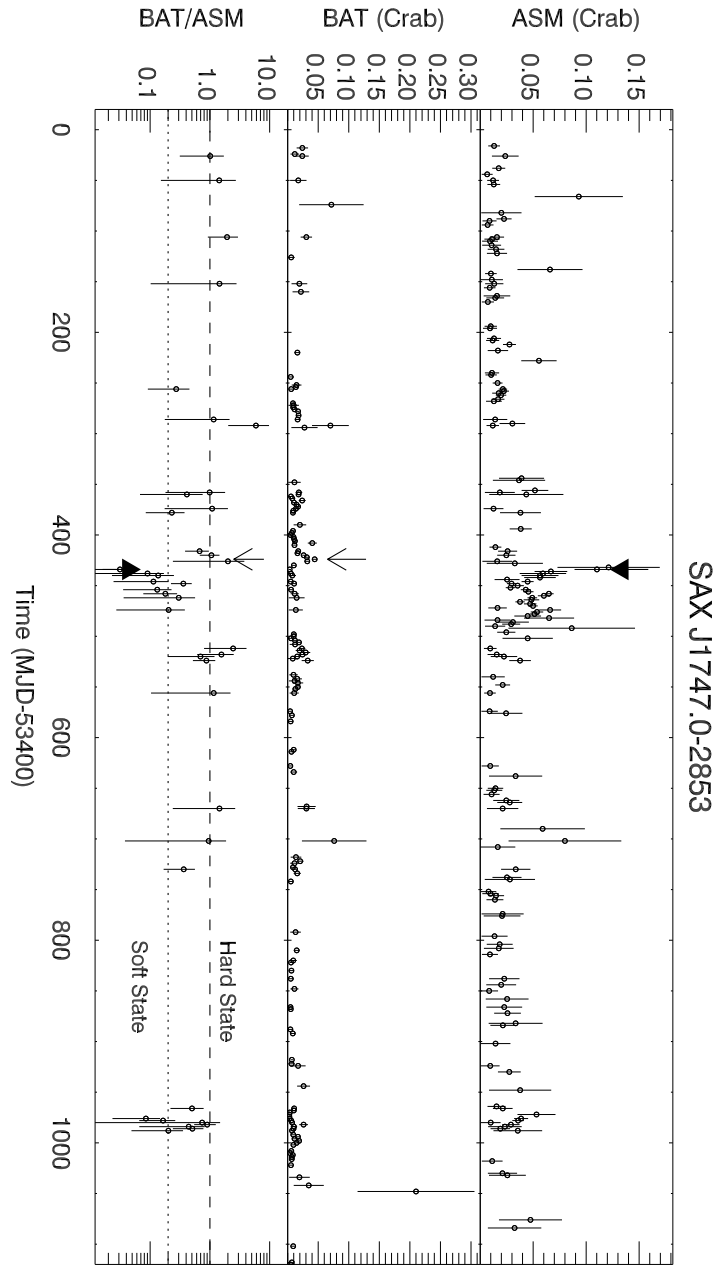}
\caption{X-ray monitoring observations of SAX J1747.0-2853 in 2--12 keV with the ASM and 15--50 keV with the BAT. }
\end{figure*}
\clearpage

\begin{figure*}
\epsscale{0.6}
\plotone{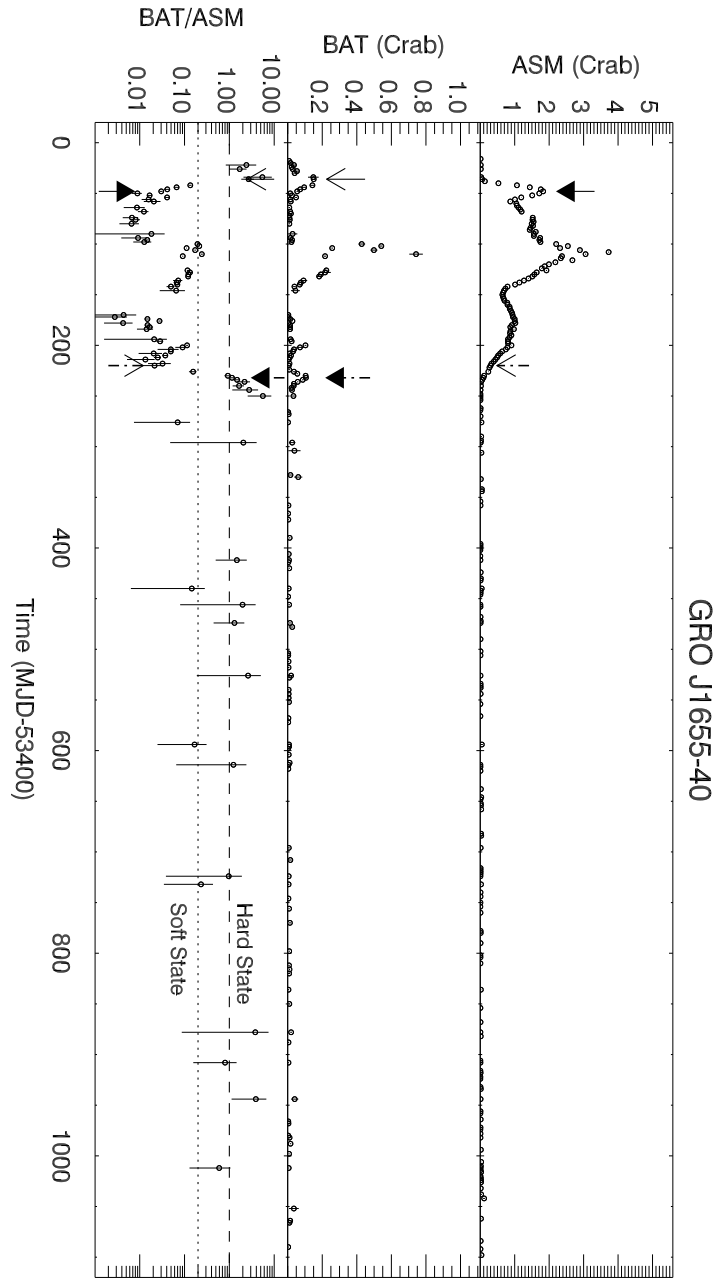}
\caption{X-ray monitoring observations of GRO J1655-40 in 2--12 keV with the ASM and 15--50 keV with the BAT. }
\end{figure*}
\clearpage

\begin{figure*}
\epsscale{0.6}
\plotone{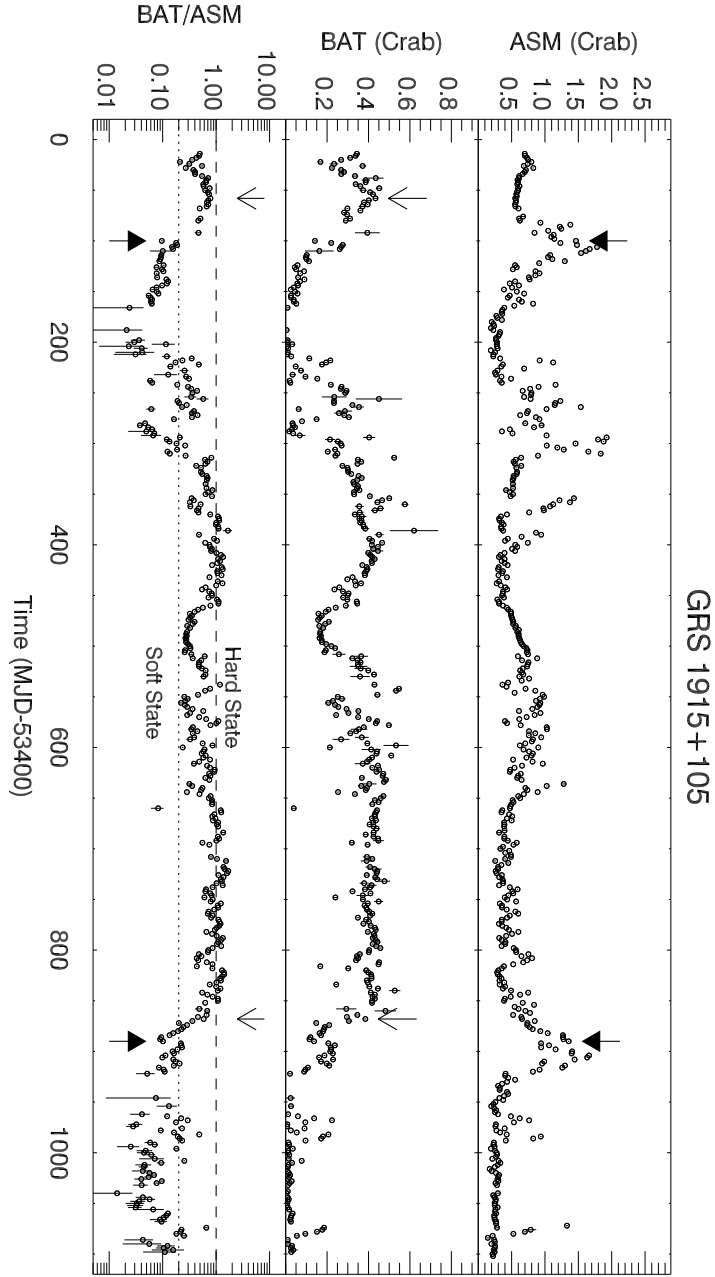}
\caption{X-ray monitoring observations of GRS 1915+105 in 2--12 keV with the ASM and 15--50 keV with the BAT. The hardness ratios for the hard state are lower than other black hole binaries. In order to include the transition from the hard state to the soft state around Day 60 as the second transition sample, we lower the hardness ratio threshold for the hard state to 0.6. }
\end{figure*}
\clearpage

\begin{figure*}
\epsscale{0.6}
\plotone{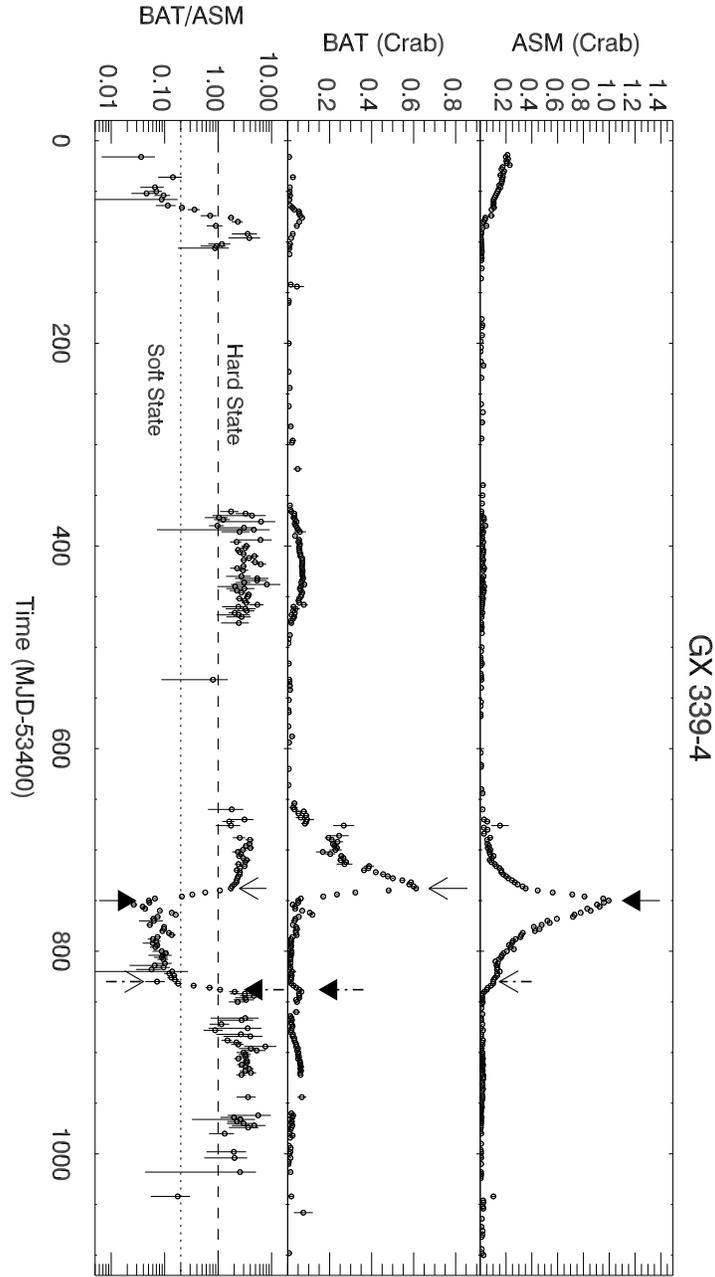}
\caption{X-ray monitoring observations of GX 339-4 in 2--12 keV with the ASM and 15--50 keV with the BAT.}
\end{figure*}
\clearpage

\begin{figure*}
\epsscale{0.6}
\plotone{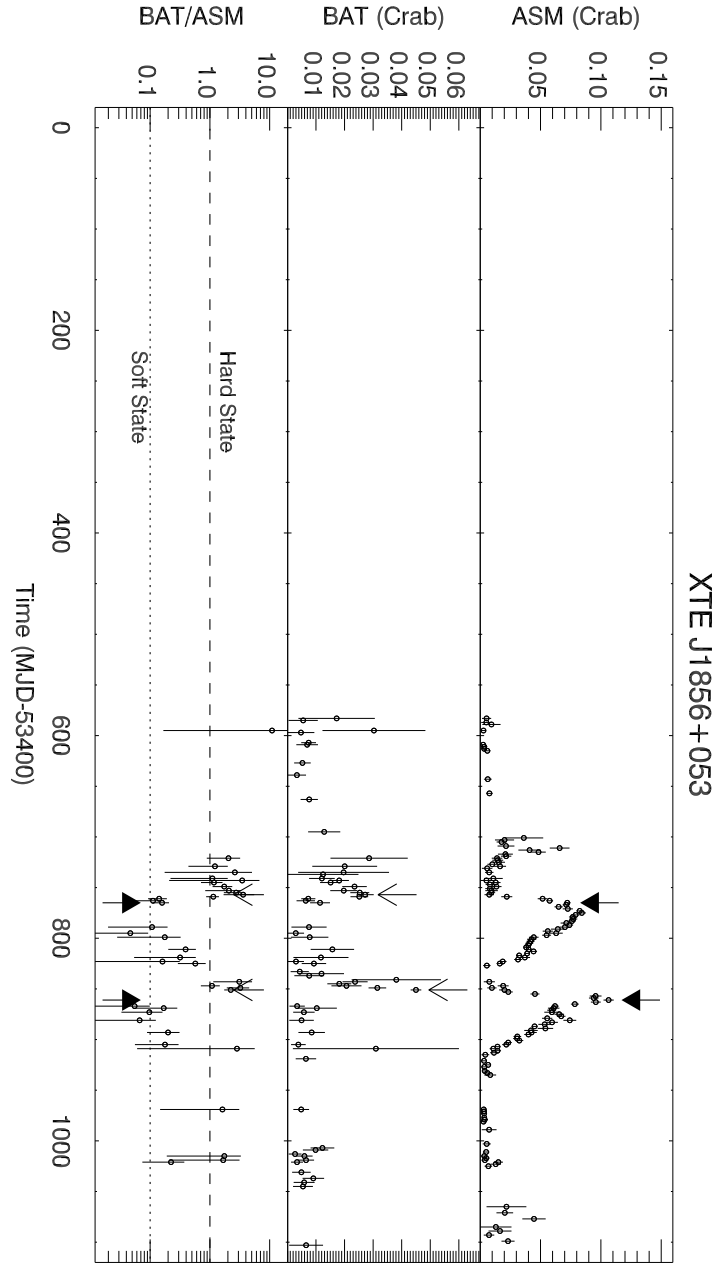}
\caption{X-ray monitoring observations of XTE J1856+053 in 2--12 keV with the ASM and 15--50 keV with the BAT.}
\end{figure*}
\clearpage

\begin{figure*}
\epsscale{0.6}
\plotone{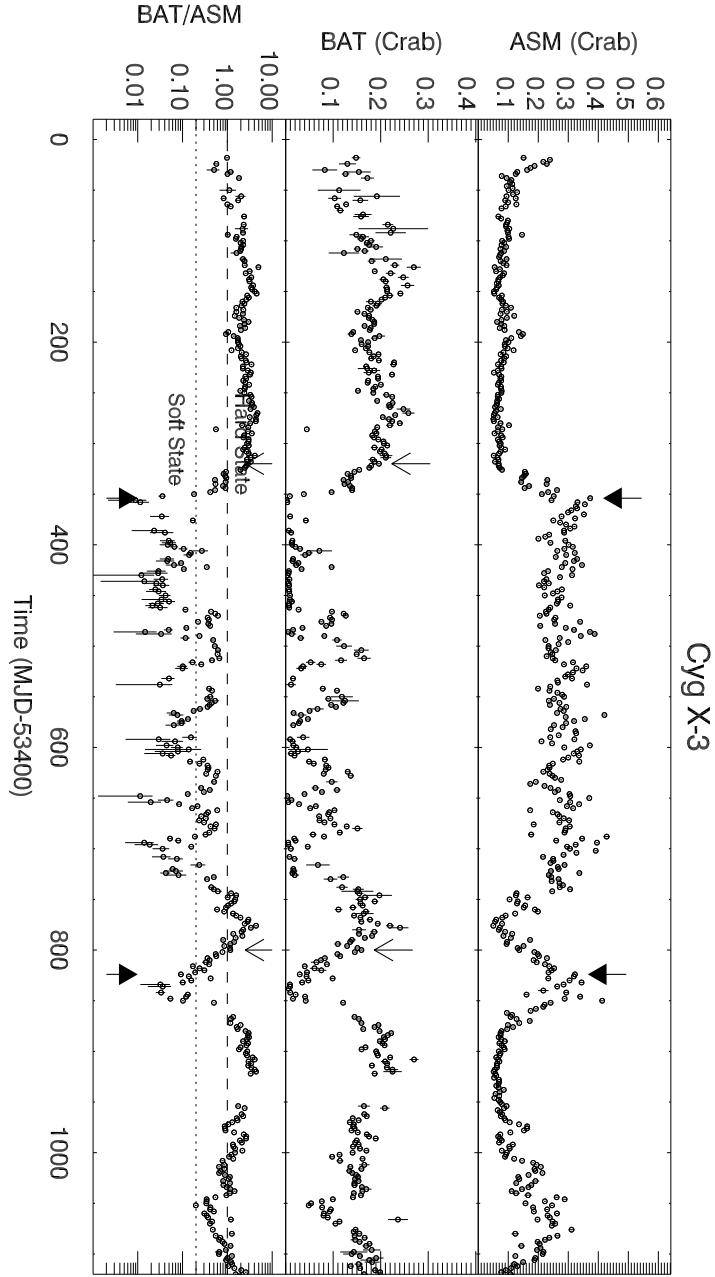}
\caption{X-ray monitoring observations of Cyg X-3 in 2--12 keV with the ASM and 15--50 keV with the BAT. The thresholds for spectral states in black hole systems were used. }
\end{figure*}
\clearpage

\begin{figure*}
\epsscale{0.6}
\plotone{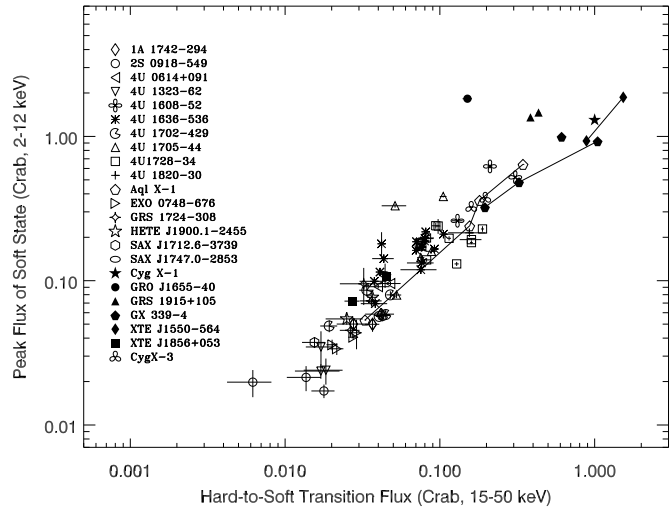}
\caption{The observed BAT fluxes when the H-S transitions occurred and the corresponding ASM peak fluxes of the following soft states. Data connected with a straight line were taken from previous studies of Aql X-1 \citep{yd07}, GX 339-4 \citep{yu07}, and XTE J1550-564 \citep{yu04} with  pointed observations, respectively.}
\end{figure*}
\clearpage

\epsscale{0.6}

\begin{figure*}
\epsscale{0.6}
\plotone{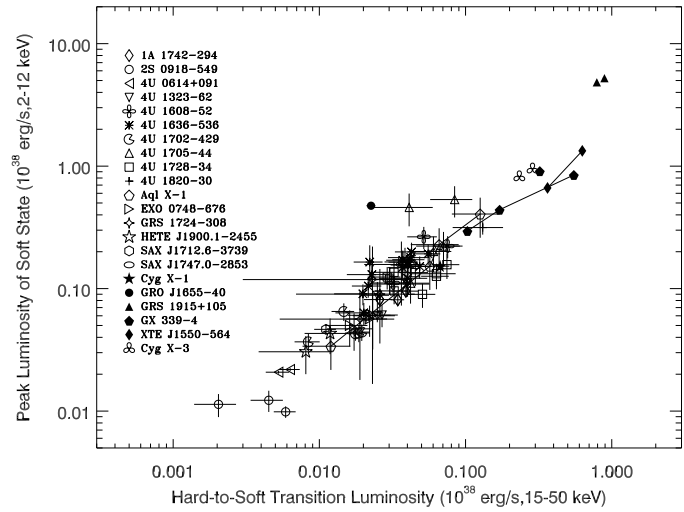}
\caption{The correlation between the transition luminosity (15--50 keV, ergs/s) and the corresponding peak luminosity of the following soft state (2--12 keV, ergs/s). The luminosities were estimated based on the X-ray energy spectrum of the Crab (Kirsth et al. 2005). Data connected with a solid line were from previous studies with pointed observations of single sources, see Fig. 22.}
\end{figure*}
\clearpage

\begin{figure*}
\epsscale{0.6}
\plotone{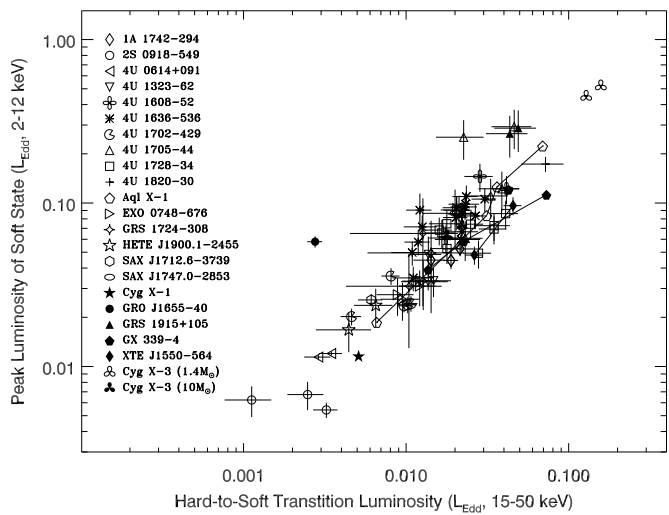}
\caption{The correlation between the transition luminosity (15--50 keV) and the peak luminosity of the following soft state (2--12 keV) in Eddington units. Source distances and masses used are listed in Table 1. Data point for Cygnus X-1 is based on CGRO/BATSE and RXTE/ASM observations in 1996 (Zhang et al. 1997). Data connected by a solid line were from previous single source studies, as those in Fig.~22. Notice that 2S 0918-549, Cyg X-1, GRS 1915+105, 4U1820-30, and 4U 1705-44 are at both ends of the correlation with rather accurate distance estimates, indicating that the luminosity of the H-S transition spans by two orders of magnitude. }
\end{figure*}
\clearpage

\begin{figure*}
\epsscale{0.6}
\plotone{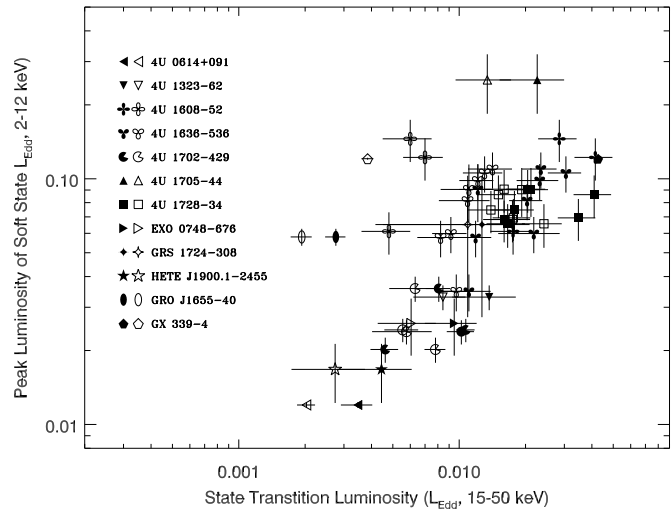}
\caption{Comparison between the luminosities of the hard-to-soft and the soft-to-hard transitions associated with the same outbursts or flares. Filled and unfilled symbols represent those of the hard-to-soft transitions and the soft-to-hard transitions, respectively. }
\end{figure*}
\clearpage

\begin{figure*}
\epsscale{0.6}
\plotone{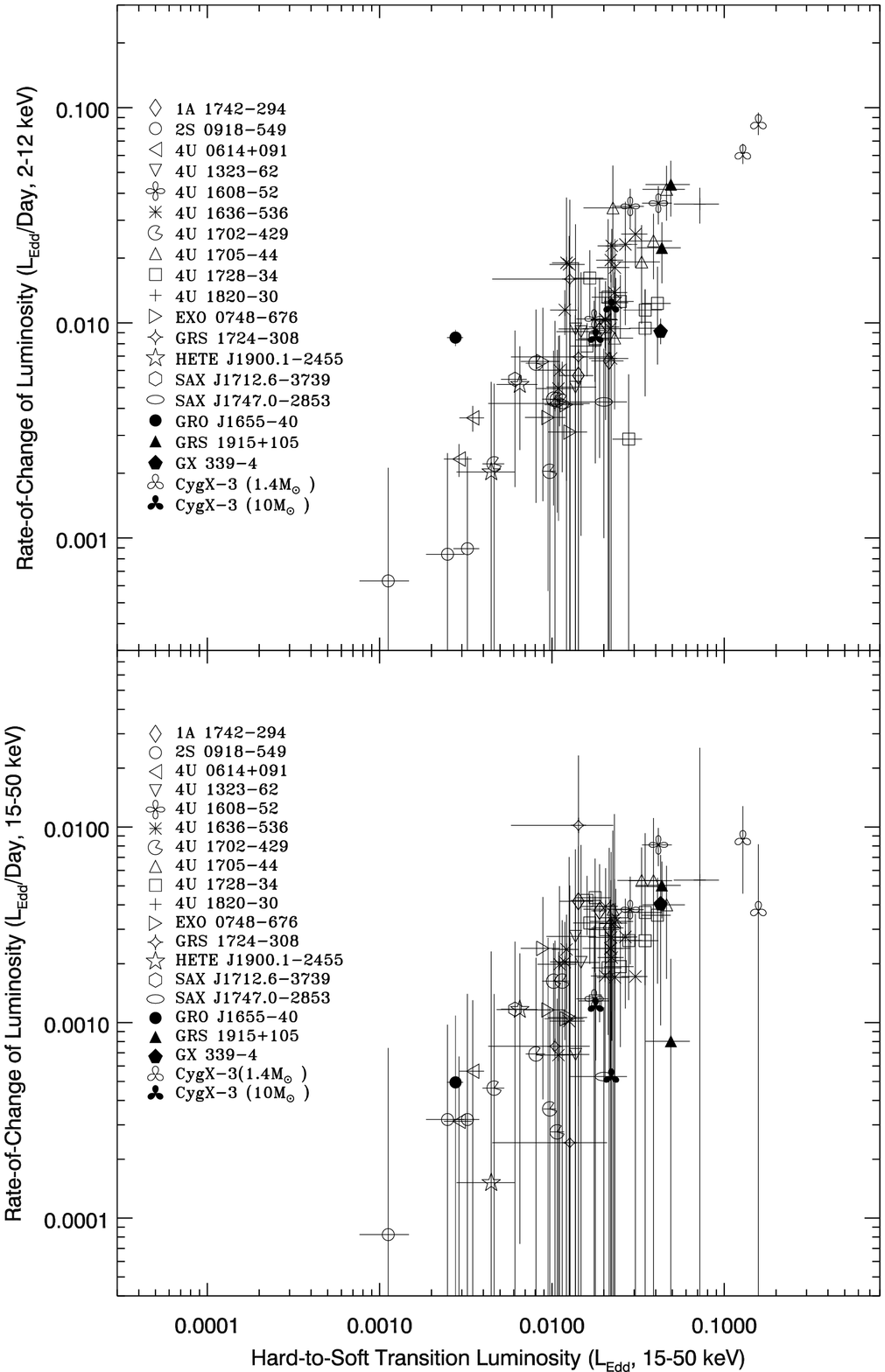}
\caption{The correlation between the luminosity of the hard-to-soft transition and the maximum rate-of-increase of the X-ray luminosity around the hard-to-soft transition.}
\end{figure*}
\clearpage

\begin{figure*}
\epsscale{0.6}
\plotone{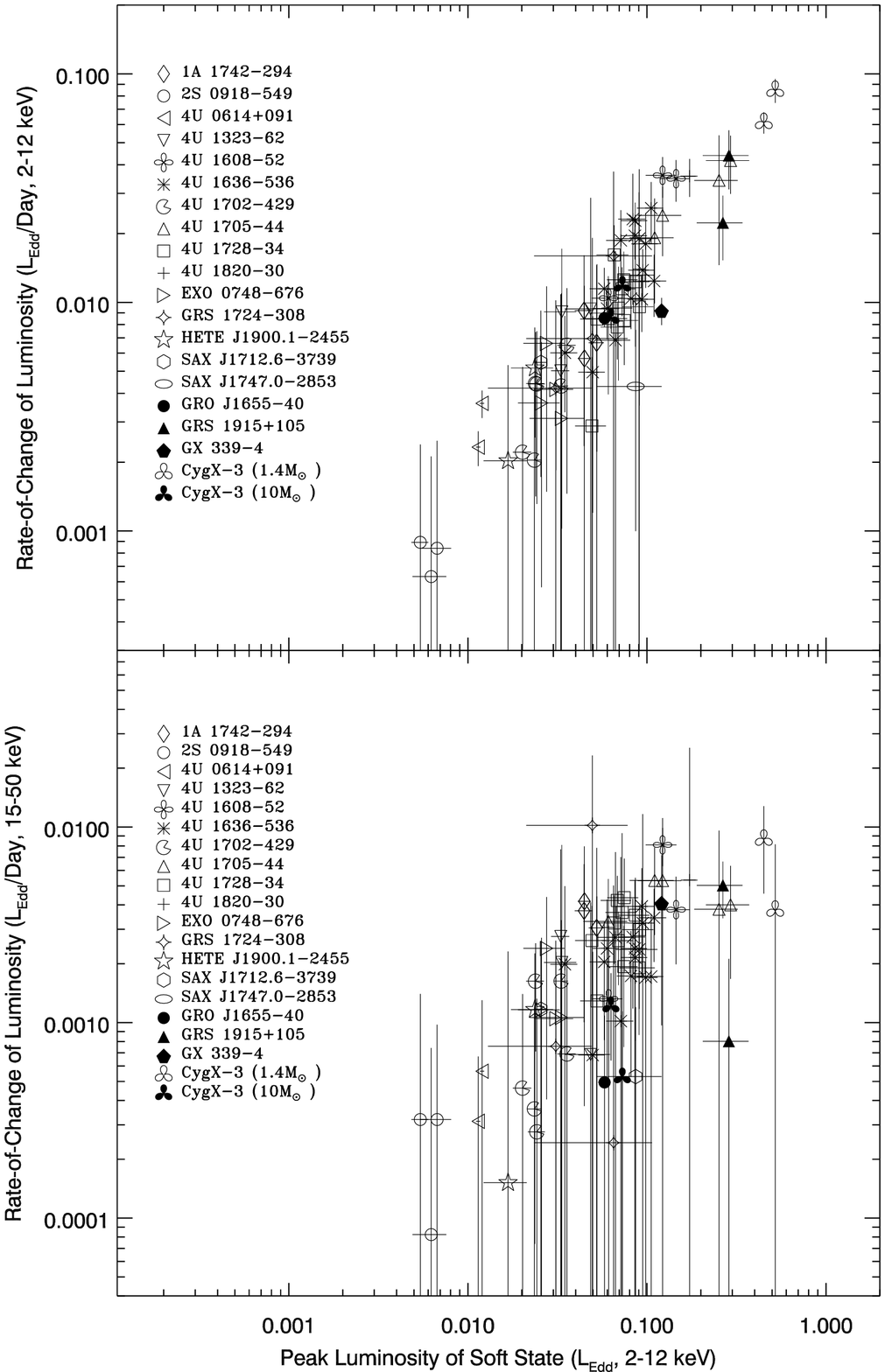}
\caption{The correlation between the peak luminosity of the soft state and the maximum rate-of-increase of the X-ray luminosity around the hard-to-soft transition.}
\end{figure*}
\clearpage

\begin{figure*}
\epsscale{1.0}
\plotone{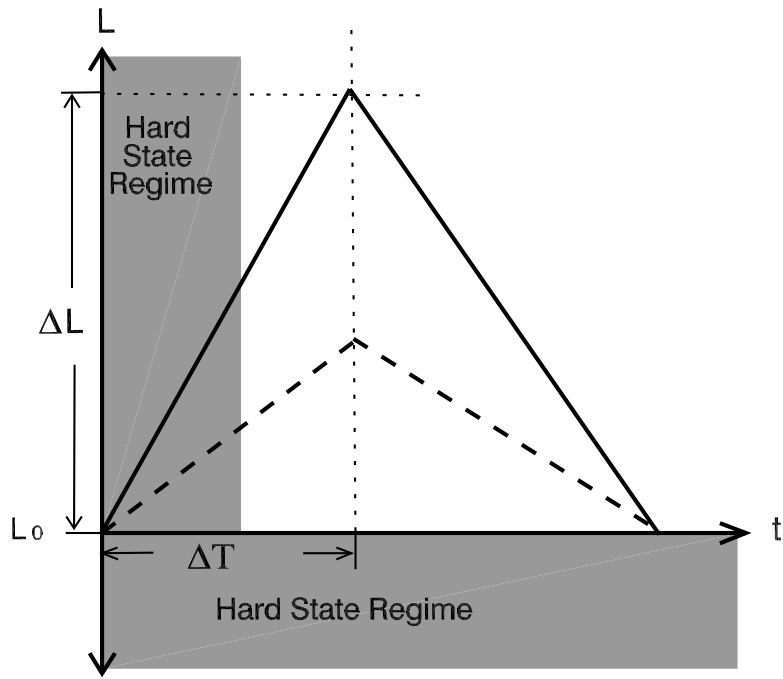}
\caption{A schematic picture of the regimes of the hard state. Two assumed transient outbursts of different peak luminosities are shown as solid curve and dashed curve, respectively. When a source is under stationary accretion, spectral transitions between the hard state and the soft state occurs at a nearly a constant luminosity $\rm L_{0}$, as expected from the framework that the mass accretion rate determines spectral states. When a source is undergoing an outburst or flare, the hard-to-soft transition occurs at a luminosity above $\rm L_{0}$. The additional luminosity roughly proportional to $\frac{\Delta L}{\Delta T}$. The soft-to-hard transitions are expected to occur around $\rm L_{0}$, but effect of non-stationary accretion may exist as well (see e.g., Smith et al. 2007). }
\end{figure*}
\clearpage

\begin{deluxetable}{cccccc}
\tabletypesize{\scriptsize}
\tablewidth{1pt}
\tablenum{1}
\tablecaption{List of the sources with H-S transitions identified and parameters used*}
\tablewidth{0pt}
\tablehead{\colhead{Source}    & \colhead{Distance}   &
           \colhead{Mass }      & \colhead{H2S}&
           \colhead{S2H}   & References                     \\
                                			 &\colhead{($\rm kpc$)}     &
           \colhead{($M_\sun$)}   }
\startdata
1A 1742-294 &  8.0 &1.4 &3 &0	& \citet{be06} \\
2S 0918-549 & 4.1-5.4 & 1.4 & 3 &0	& \citet{in05}\\
4U 0614+091  & 3 & 1.4 &2 &1 	& \citet{br92}\\
4U 1323-62 & 10 &1.4 &3 &1	& \citet{pa89} \\
4U 1608-52 & 4.1$\pm$0.4 & 1.4 &3 &3 & \citet{ga06}\\
4U 1636-536 & 6$\pm$0.5 & 1.4 &16 &8 &  \citet{ga06}\\
4U 1702-429 & 5.46$\pm$0.19 &1.4 &6 &4 & \citet{ga06}\\
4U 1705-44 & 7.4$^{+0.8}_{-1.1}$ &1.4 &5 &1 & \citet{ht95}\\
4U 1728-34 & 5.2$\pm$0.5 & 1.4 &11 &7	& \citet{ga06}\\
4U 1820-30 & 7.6$\pm$0.4 & 1.4 &1 &0 & \citet{ku03}\\
EXO 0748-676 & 7.4$\pm$0.9 & 1.4 &4 &1 & \citet{ga06,wo05}\\
GRS 1724-308 &7$\pm$2 &1.4 &3 &1 & \citet{ga06}\\
HETE J1900.1-2455 & 4.7$\pm$0.6 & 1.4 &2 &1 & \citet{ga06}\\
SAX J1712.6-3739 &7 &1.4 &1 &0 & \citet{co01}\\
SAX J1747.0-2853 &7.5$\pm$1.3 &1.4 &1 &0 & \citet{we04}\\
GRO 1655-40 & 3.2 & 6.3$\pm$0.5 &1 &1 & \citet{gr01,hr95}\\
GRS 1915+105 & 11.2-12.5 & 14$\pm$4 &2 &0 & \citet{gre01,fe99,mr94}\\
GX 339-4 & $\geq$5.6 &$\geq$5.8$\pm$0.5 &1 &1 & \citet{hy03,sh01}\\
XTE J1856+053 &unknown &unknown &2 &0 & \\
Cyg X-3 &10 &unknown &2 &0 & \citet{di83}\\
\enddata

\tablenotetext{*}{For neutron stars, no accurate mass measurement is known. So 1.4 solar masses were used. For GX 339-4, only lower limits of the distance and the black hole mass, 5.6 kpc and 5.8 solar masses, are known. These values were used as the actual distance and mass.}
\end{deluxetable}


\clearpage

\begin{thebibliography}{}
\bibitem[B{\'e}langer et al.(2006)]{be06} B{\'e}langer, G., et al.\ 2006, \apj, 636, 275
\bibitem[Belloni et al.(2002)]{be02} Belloni, T., Colombo, A.~P., Homan, J., Campana, S., \& van der Klis, M.\ 2002, \aap, 390, 199 
\bibitem[Bouchacourt et al.(1984)]{bo84} Bouchacourt, P., Chambon, G., Niel, M., Refloch, A., Estulin, I.~V., Kuznetsov, A.~V., \& Melioranskii, A.~S.\ 1984, \apjl, 285, L67 
\bibitem[Brandt et al.(1992)]{br92} Brandt, S. et al. 1992, {\aap}, 262L, 15
\bibitem[Brocksopp et al.(2001)]{br01} Brocksopp, C., Jonker, P.~G., Fender, R.~P., Groot, P.~J., van der Klis, M., \& Tingay, S.~J.\ 2001, \mnras, 323, 517 
\bibitem[Brocksopp et al.(2002)]{br02} Brocksopp, C., et al.\ 2002, \mnras, 331, 765 
\bibitem[Brocksopp et al.(2004)]{br04} Brocksopp, C., Bandyopadhyay, R.~M., \& Fender, R.~P.\ 2004, New Astronomy, 9, 249 
\bibitem[Brocksopp et al.(2006)]{brocksopp06} Brocksopp, C. et al. 2006, \mnras, 365, 1203
\bibitem[Cocchi et al.(2001)]{co01} Cocchi, M., Bazzano, A., Natalucci, L., Ubertini, P., Heise, J., Kuulkers, E., Cornelisse, R., \& in't Zand, J.~J.~M.\ 2001, \aap, 378, L37 
\bibitem[Corbel et al.(2003)]{corbel03} Corbel, S., Nowak, M.~A., Fender, R.~P., Tzioumis, A.~K., \& Markoff, S.\ 2003, \aap, 400, 1007 
\bibitem[Dickey(1983)]{di83} Dickey, J.~M.\ 1983, \apjl, 273, L71
\bibitem[Done et al.(2007)]{done07} Done, C., Gierli{\'n}ski, M., \& Kubota, A.\ 2007, \aapr, 15, 1 
\bibitem[Esin et al.(1997)]{esin97} Esin, A.~A., McClintock, J.~E., \& Narayan, R.\ 1997, \apj, 489, 865 
\bibitem[Fender et al.(1999)]{fe99} Fender R. P. et al. 1999, {\mnras}, 304, 865
\bibitem[Galloway et al.(2006)]{ga06} Galloway, D.K. et al. 2006, astro-ph/0608259
\bibitem[Gladstone et al.(2007)]{gladstone07} Gladstone, J., Done, C., \& Gierli{\'n}ski, M.\ 2007, \mnras, 378, 13 
\bibitem[Greene et al.(2001)]{gr01} {{Greene}, J. and {Bailyn}, C.~D. and {Orosz}, J.~A.} 2001, {\apj}, 554, 1290
\bibitem[Greiner et al.(2001)]{gre01}{{Greiner}, J. and {Cuby}, J.~G. and {McCaughrean}, M.~J.} 2001, {\nat}, 414, 522
\bibitem[Grimm et al.(2002)]{grimm02} Grimm, H.-J., Gilfanov, M., \& Sunyaev, R.\ 2002, \aap, 391, 923 
\bibitem[Haberl \& Titarchuk(1995)]{ht95} {{Haberl}, F. and {Titarchuk}, L.} 1995, \aap, 299, 414
\bibitem[Hjellming \& Rupen(1995)]{hr95} {{Hjellming}, R.~M. and {Rupen}, M.~P.} 1995, {\nat}, 375, 464
\bibitem[Homan et al. (2001)]{homan01} Homan, J., Wijnands, R., van der Klis, M., Belloni, T., van Paradijs, J., Klein-Wolt, M., Fender, R., \& M{\'e}ndez, M.\ 2001, \apjs, 132, 377 
\bibitem[Homan \& Belloni(2005)]{homan05} Homan, J., \& Belloni, T.  2005, \apss, 300, 107
\bibitem[in't Zand et al.(2005)]{in05} in't Zand, J. J. M. et al. 2005, {\aap}, 441, 675
\bibitem[Hynes et al.(2003)]{hy03} {{Hynes}, R.~I. and {Steeghs}, D. and {Casares}, J. and {Charles}, P.~A. and {O'Brien}, K.}, 2003, \apj, 583, 95
\bibitem[Kaaret et al.(1998)]{kaaret98} Kaaret, P., Yu, W., Ford, E.~C., \& Zhang, S.~N.\ 1998, \apjl, 497, L93 
\bibitem[Kaaret et al. (2006)]{kaaret06} Kaaret, P., Simet, M. G. \& Lang, C. C. 2006, \apj, 646,174
\bibitem[Kirsch et al.(2005)]{kirsch05} Kirsch, M. G., et al. 2005, SPIE, 5898, 22
\bibitem[Kong et al. (2004)]{kong04} Kong, A. K. H., Di Stefano, R., \& Yuan, F. 2004, 617, L49
\bibitem[Kuulkers et al.(2003)]{ku03} {{Kuulkers}, E. and {den Hartog}, P.~R. and {in't Zand}, J.~J.~M. and {Verbunt}, F.~W.~M. and {Harris}, W.~E. and {Cocchi}, M.} 2003 \aap, 399, 663
\bibitem[Liu et al.(2005)]{liu05} Liu, B.~F., Meyer, F., \& Meyer-Hofmeister, E.\ 2005, \aap, 442, 555 
\bibitem[Maccarone(2003)]{maccarone03} Maccarone, T.  2003, \aap, 399, 1151
\bibitem[Maccarone \& Coppi(2003)]{mc03} Maccarone, T. J., \& P. S. Coppi  2003, \mnras, 338, 189
\bibitem[Massey et al.(1995)]{massey95} Massey, P., Johnson, K.~E., \& Degioia-Eastwood, K.\ 1995, \apj, 454, 151 
\bibitem[Meyer-Hofmeister et al.(2004)]{meyer04} Meyer-Hofmeister, E., Liu, B. F., \& Meyer, F.  2005, \aap, 432, 181
\bibitem[Meyer-Hofmeister et al.(2005)]{meyer05} Meyer-Hofmeister, E., Liu, B.~F., \& Meyer, F.\ 2005, \aap, 432, 181 
\bibitem[Mirabel \& Rodriguez(1994)]{mr94}  {{Mirabel}, I.~F. and {Rodriguez}, L.~F.} 1994, {\nat}, 371, 46
\bibitem[Miyamoto et al.(1995)]{miyamoto95} Miyamoto, S., et al.  1995, \apj, 442, L13
\bibitem[Narayan \& Yi(1994)]{na94} Narayan, R., \& Yi, I.\ 1994, \apjl, 428, L13 
\bibitem[Narayan \& Yi(1995)]{na95} Narayan, R., \& Yi, I.\ 1995, \apj, 452, 710 
\bibitem[Nowak(1995)]{nowak95} Nowak, M.~A.\ 1995, \pasp, 107, 1207
\bibitem[Orosz et al.(2002)]{or02} Orosz, J.~A., et al.\ 2002, \apj, 568, 845
\bibitem[Ortolani et al.(1994)]{ortolani94} {{Ortolani}, S. and {Barbuy}, B. and {Bica}, E.}  1994, \aaps, 108, 653
\bibitem[Parmar et al.(1989)]{pa89} Parmar, A. N. et al. 1989, \apj, 338, 1024
\bibitem[Ramadevi \& Seetha(2007)]{ra07} Ramadevi, M.~C., \& Seetha, S.\ 2007, \mnras, 378, 182
\bibitem[Remillard \& McClintock (2006)]{re06} Remillard, R. A., \& McClintock, J. E.\ 2006, \araa, 44, 49
\bibitem[Rodriguez et al.(2006)]{ro06} Rodriguez, J., Shaw, S.~E., \& Corbel, S.\ 2006, \aap, 451, 1045 
\bibitem[Rodriguez et al.(2007)]{ro07} Rodriguez, J., Bel, M.~C., Tomsick, J.~A., Corbel, S., Brocksopp, C., Paizis, A., Shaw, S.~E., 
\& Bodaghee, A.\ 2007, \apjl, 655, L97
\bibitem[Russell et al.(2007)]{ru07} Russell, D.~M., Maccarone, T.~J., K{\"o}rding, E.~G., \& Homan, J.\ 2007, \mnras, 379, 1401 
\bibitem[Rutledge et al.(2001)]{ru01} {{Rutledge}, R.~E. and {Bildsten}, L. and {Brown}, E.~F. and {Pavlov}, G.~G. and {Zavlin}, V.~E.} 2001, \apj, 559, 1054
\bibitem[Shahbaz et al.(2001)]{sh01} T. Shahbaz et al. 2001, \aap, 376, L17-L21
\bibitem[Smith et al.(2002)]{smith02} Smith, D. M., Heindl, W. A., \& Swank, J. H.  2002, \apj, 569, 362
\bibitem[Smith et al.(2007)]{smith07} Smith, D.~M., Dawson, D.~M., \& Swank, J.~H.\ 2007, \apj, 669, 1138
\bibitem[Soria et al.(2007)]{soria07} Soria, R., Baldi, A., Risaliti, G., Fabbiano, G., King, A., La Parola, V., \& Zezas, A.\ 2007, \mnras, 379, 1313 
\bibitem[van der Klis(2001)]{klis01} van der Klis, M.\ 2001, \apj, 561, 943 
\bibitem[Werner et al.(2004)]{we04} Werner, N., et al.\ 2004, \aap, 416, 311 
\bibitem[Wilms et al.(2006)]{wilms06} Wilms, J., Nowak, M.~A., Pottschmidt, K., Pooley, G.~G., \& Fritz, S.\ 2006, \aap, 447, 245 
\bibitem[Wolff et al.(2005)]{wo05} {{Wolff}, M.~T. and {Becker}, P.~A. and {Ray}, P.~S. and {Wood}, K.~S.} 2005, {\apj}, 632, 1099
\bibitem[Yu et al.(2003)]{yu03} Yu, W., Klein-Wolt, M., Fender, R., \& van der Klis, M.\ 2003, \apjl, 589, L33 
\bibitem[Yu, van der Klis \& Fender (2004)]{yu04} Yu, W., van der Klis, M., \& Fender, R.  2004, \apj, 611, L121
\bibitem[Yu \& Dolence(2007)]{yd07} Yu, W., \& Dolence, J.  2007, \apj, 667, 1043
\bibitem[Yu et al.(2007)]{yu07} Yu,W., Lamb, F. K., Fender, R. \& van der Klis, M.  2007, \apj, 663, 1309
\bibitem[Zdziarski et al.(2002)]{zd02} Zdziarski, A.~A., Poutanen, J., Paciesas, W.~S., \& Wen, L.\ 2002, \apj, 578, 357 
\bibitem[Zdziarski et al.(2004)]{zd04} Zdziarski, Andrzej A., et al.  2004, \mnras, 351, 791
\bibitem[Zhang et al.(1997)]{zhang97} Zhang S. N. et al.  1997, \apj, 477, L95
\end{thebibliography}
\end{document}